\documentclass[useAMS,usenatbib]{mnras}
\usepackage{graphicx}
\usepackage{amsmath, amssymb}
\usepackage[T1]{fontenc}
\usepackage[usenames]{color}
\usepackage[dvipsnames]{xcolor}
\usepackage{geometry}
\usepackage{bm}
\usepackage[capitalize]{cleveref}
\usepackage{physics}
\usepackage{acro}
\usepackage{caption}
\usepackage{subcaption}
\usepackage{stfloats}
\usepackage{xspace}
\usepackage{booktabs}
\usepackage{tabularx}
\usepackage{multicol}
\usepackage{url}
\usepackage{hyperref}
\usepackage{orcidlink}
\usepackage[group-separator={,}]{siunitx}

\hypersetup{breaklinks=true}
\newcommand{\zobs}{z_{\rm CMB}}
\newcommand{\zpred}{z_{\rm pred}}
\newcommand{\ztrue}{z_{\rm true}}
\newcommand{\zCMB}{z_{\rm CMB}}
\newcommand{\zcosmo}{z_{\rm cosmo}}
\newcommand{\zpec}{z_{\rm pec}}

\newcommand{\Vpec}{V_{\rm pec}}
\newcommand{\Vext}{\bm{V}_{\rm ext}}

\newcommand{\Mpch}{\ensuremath{h^{-1}\,\mathrm{Mpc}}}
\newcommand{\kmsec}{\ensuremath{\mathrm{km}\,\mathrm{s}^{-1}}}
\newcommand{\kmsecMpc}{\ensuremath{\mathrm{km}\,\mathrm{s}^{-1}\,\mathrm{Mpc}^{-1}}}

\newcommand{\mpred}{m_{\rm pred}}
\newcommand{\mobs}{m_{\rm obs}}
\newcommand{\mtrue}{m_{\rm true}}
\newcommand{\etaobs}{\eta_{\rm obs}}
\newcommand{\etatrue}{\eta_{\rm true}}

\newcommand{\sint}{\sigma_{\rm int}}
\newcommand{\aTFR}{a_{\rm TFR}}
\newcommand{\bTFR}{b_{\rm TFR}}
\newcommand{\cTFR}{c_{\rm TFR}}

\newcommand{\MSN}{M_{\rm SN}}

\newcommand{\Om}{\Omega_{\rm m}}
\newcommand{\SFI}{SFI\texttt{++}\xspace}
\newcommand{\TWOMPP}{2M\texttt{++}\xspace}
\newcommand{\PP}{Pantheon\texttt{+}\xspace}
\newcommand{\PPL}{Pantheon\texttt{+}Lane\xspace}
\newcommand{\ebv}{\ensuremath{E(B\!-\!V)}}

\defcitealias{Carrick_2015}{C15}
\defcitealias{VF_olympics}{S25}
\defcitealias{Boubel_2025}{B25}

\DeclareAcronym{CMB}{short = CMB, long  = cosmic microwave background}
\DeclareAcronym{TFR}{short = TFR, long  = Tully--Fisher relation}
\DeclareAcronym{LCDM}{short = $\Lambda$CDM, long  = $\Lambda$-cold dark matter}
\DeclareAcronym{CF4}{short = CF4, long  = CosmicFlows-4}
\DeclareAcronym{HMC}{short = HMC, long  = Hamiltonian Monte Carlo}
\DeclareAcronym{NUTS}{short = NUTS, long  = No-U-Turn Sampler}
\DeclareAcronym{SDSS}{short = SDSS, long = Sloan Digital Sky Survey}
\DeclareAcronym{WISE}{short = WISE, long = Wide-field Infrared Survey Explorer}
\DeclareAcronym{MCMC}{short = MCMC, long = Markov Chain Monte Carlo}
\DeclareAcronym{CP}{short = CP, long = cosmological principle}
\DeclareAcronym{SN}{short = SN, long  = supernova, short-plural = e, long-plural  = e}

\title[No evidence for $H_0$ anisotropy from TF or SN distances]{No evidence for local $H_0$ anisotropy from Tully--Fisher or supernova distances}

\author[R. Stiskalek, H. Desmond and G. Lavaux]{Richard Stiskalek$^{1}$\thanks{\href{mailto:richard.stiskalek@physics.ox.ac.uk}{richard.stiskalek@physics.ox.ac.uk}}\orcidlink{0000-0002-0986-314X},
Harry Desmond$^{2}$\orcidlink{0000-0003-0685-9791}
and Guilhem Lavaux$^{3}$\orcidlink{0000-0003-0143-8891}
\\
$^{1}$Astrophysics, University of Oxford, Denys Wilkinson Building, Keble Road, Oxford, OX1 3RH, UK\\
$^{2}$Institute of Cosmology \& Gravitation, University of Portsmouth, Dennis Sciama Building, Portsmouth, PO1 3FX, UK\\
$^{3}$CNRS \& Sorbonne Universit\'e, Institut d'Astrophysique de Paris (IAP), UMR 7095, 98 bis bd Arago, F-75014 Paris, France\\
}

\date{Accepted XXX. Received YYY; in original form ZZZ}
\pubyear{2025}

\begin{document}\label{firstpage}
\pagerange{\pageref{firstpage}--\pageref{lastpage}}
\maketitle

\begin{abstract}
Claims of local ($z \lesssim 0.05$) anisotropy in the Hubble constant have been made based on direct distance tracers such as Tully--Fisher galaxies and Type Ia supernovae. We revisit these using the CosmicFlows-4 Tully--Fisher W1 subsample, 2MTF and \SFI\ Tully--Fisher catalogues, and the \PP\ supernova compilation (all restricted to $z < 0.05$), including a dipole in either the Tully--Fisher zero-point or the standardised supernova absolute magnitude. Our forward-modelling framework jointly calibrates the distance relation, marginalises over distances, and accounts for peculiar velocities using a linear-theory reconstruction. We compare the anisotropic and isotropic model using the Bayesian evidence. In the CosmicFlows-4 sample, we infer a zero-point dipole of amplitude $0.087 \pm 0.019$ mag, or $4.1\pm0.9$ per cent when expressed as a dipole in the Hubble parameter. This is consistent with previous estimates but at higher significance: model comparison yields odds of $877\!:\!1$ in favour of including the zero-point dipole. In \PP\ we infer zero-point dipole amplitude of $0.049 \pm 0.013$ mag, or $2.3\pm 0.6$ per cent when expressed as a dipole in the Hubble parameter. However, by allowing for a radially varying velocity dipole, we show that the anisotropic zero-point model captures local flow features (or possibly systematics) in the data rather than an actual linearly growing effective bulk flow caused by anisotropy in the zero-point or expansion rate. Crucially, inferring a more general bulk flow curve we find results fully consistent with expectations from the standard cosmological model.
\end{abstract}



\section{Introduction}\label{sec:intro}

The \ac{CP}---that the Universe is isotropic and homogeneous on sufficiently large scales---plays a foundational role in modern cosmology. Combined with the assumption of General Relativity as the theory of gravity, it leads to the Friedmann--Robertson--Walker metric which underpins the concordance model of cosmology, \ac{LCDM}. It is therefore crucial to test if the \ac{CP} is satisfied and hence fit for the purpose of providing such a bedrock. Recent years have seen increased scrutiny of the \ac{CP}, for example in the \acl{CMB} (\ac{CMB};~\citealt{CMB_anisotropy,Sravan}), Type Ia supernovae~\citep{Basheer,Hu,SN_1,SN_2,Sah,Conville,Krishnan,Zhai,Rahman_2022,Cowell_2023,Sorrenti_2025}, direct distance tracers~\citep{Watkins_2023,Boubel_2025}, galaxy clusters~\citep{Migkas_2021,Pandya_2024} and distant radio galaxies, quasars and gamma ray bursts~\citep{EB_1,EB_2,colloquium,Luongo_2022} (for a review, see~\citealt{CP_review}).
Moreover, increasing attention has been directed toward defining ``expansion'' in metrics beyond Friedmann--Robertson--Walker to enable model-independent tests~\citep{Maartens_2024,Kalbouneh_2024,Kalbouneh_2025,Sarma_2025}.
Any evidence for anisotropy must be carefully validated before violation of so fundamental a principle as the \ac{CP} can be concluded.

One powerful test of the \ac{CP} involves examining the \emph{local} expansion rate $H_0$ in different directions, which can be achieved by using direct distance tracers such as \ac{TFR} galaxies and Type Ia \acp{SN}. This is doubly important as it may also be relevant for the Hubble tension, which is derived assuming isotropic expansion~\citep{Riess_2022,cosmoverse}. Without calibration through a lower rung of the distance ladder, an anisotropy in $H_0$ is completely degenerate with an anisotropy in the normalisation of the TFR and the absolute magnitude of \acp{SN}. While it is therefore not possible to assess an $H_0$ anisotropy directly with this method, it seems implausible that a null detection of the degenerate combination would arise from compensated anisotropies in both $H_0$ and the normalisation of the standardising relation.

In this paper, we revisit claims of anisotropy in $H_0$ by fully forward-modelling the \ac{TFR} and \ac{SN} observables at $z < 0.05$. Our starting point is the recent study of~\citet{Boubel_2025} (hereafter~\citetalias{Boubel_2025}), who conclude weak evidence for anisotropy in the \ac{CF4} W1 \ac{TFR} subsample using a partial forward model. Following~\citetalias{Boubel_2025}, we adopt the first-order deviation from isotropy in spherical harmonics, namely a dipolar modulation to the degenerate combination of the \ac{TFR} or \ac{SN} normalisation and $H_0$. Our approach jointly calibrates the distance relation, marginalizes over distances and latent parameters describing the true values of the observables, and corrects for peculiar velocities using a linear-theory reconstruction based on~\cite{Carrick_2015} (hereafter \citetalias{Carrick_2015}). We generalise the analysis of~\citetalias{Boubel_2025} to three further independent data sets to provide a more comprehensive and robust assessment of local anisotropy, namely the 2MTF and \SFI\ \ac{TFR} catalogues and the \PP\ \ac{SN} compilation (restricted to $z < 0.05$). We also, for the first time, explicitly investigate the impact of dust extinction on the results by investigating three qualitatively different maps with different priors on the extinction coefficients. This is potentially an important systematic because extinction is anisotropic across the sky.

However, we caution that although we search for a \ac{CP}-violating anomaly, this is done under the assumption of the~\citetalias{Carrick_2015} reconstruction to account for peculiar velocities, which itself assumes the \ac{CP}. In principle, the~\citetalias{Carrick_2015} reconstruction could already contain part of a cosmological dipole, which would then be subtracted in our analysis. Nevertheless, the bulk flow in~\citetalias{Carrick_2015} shows no significant deviation from \ac{LCDM} (see also e.g.~\citealt{Boruah_2019,VF_olympics}). Thus, the results presented here should be interpreted as a dipole in $H_0$ that would be superimposed on the dipole already present in the \citetalias{Carrick_2015} peculiar velocity field, which is not \ac{CP}-violating.
Nevertheless, we also consider the total bulk flow curve from both~\citetalias{Carrick_2015} and a superimposed radially varying velocity dipole to test whether the resulting flow is in tension with \ac{LCDM} expectations.

The remainder of this paper is structured as follows. In~\cref{sec:data}, we describe the data sets used in our analysis, including the \ac{CF4} \ac{TFR} sample, the 2MTF sample, the \SFI\ catalogue, and the \PP\ \ac{SN} compilation. In~\cref{sec:method}, we present our methodology, including the forward-modelling framework, the peculiar velocity modelling scheme, the Bayesian evidence calculation and the mock data procedure. \cref{sec:results} presents the results on both mock and observed data, while~\cref{sec:discussion_conclusion} presents further comparison with~\citetalias{Boubel_2025}, discusses the more general ramifications of our results, and concludes. Appendix~\ref{sec:flow_model_full} describes the flow model in full, and Appendix~\ref{sec:full_posterior} shows the complete posterior distribution for the \ac{CF4} W1 \ac{TFR} inference.

\vspace{1em}
\noindent All logarithms are base-10 unless otherwise stated. We use the notation $\mathcal{N}(x; \mu, \sigma)$ to denote a one-dimensional normal distribution with mean $\mu$ and standard deviation $\sigma$, evaluated at $x$; in higher dimensions $\mu$ is a vector and $\sigma$ is replaced by a covariance matrix. We define $h \equiv H_0 / \left(100~\kmsecMpc\right)$, where $H_0$ is the Hubble constant.


\section{Data}\label{sec:data}

To test the hypothesis of a dipole in the zero-point calibration of redshift-independent distance indicators, or, equivalently, in the Hubble constant, we analyse multiple low-redshift catalogues. Our primary data set is the \ac{TFR} subsample of the \acl{CF4} catalogue (\ac{CF4};~\citealt{Tully_2023}), restricted to photometry in the all-sky W1 band. In addition, we consider two other all-sky \ac{TFR} samples: 2MTF~\citep{Masters_2008, Hong_2019} and the~\SFI~catalogue~\citep{Springob_2007}. We also analyse the \PP\ compilation of Type Ia \ac{SN}~\citep{Scolnic_2022}, along with its reanalysis by~\cite{Lane_2024}.

Since any large-scale dipole in the zero-point is degenerate with the local peculiar velocity field, we account for peculiar velocities using the linear reconstruction of~\citetalias{Carrick_2015}. This choice, widely adopted in the literature~\citep[e.g.][]{Boruah_2019,Said_2020,Carr_2022,Boubel_2024_H0,Boubel_2025}, allows for direct comparison with recent studies, particularly~\citetalias{Boubel_2025}, and facilitates consistent treatment of large-scale flows across all samples.

\subsection{Tully--Fisher samples}

Our first method for obtaining redshift-independent distance estimates is the \acl{TFR} (\ac{TFR};~\citealt{Tully_1977}), which relates a spiral galaxy's rotational velocity to its absolute magnitude $M$. Given an observed apparent magnitude, this relation yields a distance modulus and thereby a peculiar velocity, since the observed redshift is a function of the distance and peculiar velocity. The linewidth parameter $\eta$ is defined as
\begin{equation}
    \eta \equiv \log\frac{W}{\kmsec} - 2.5,
\end{equation}
where $W$ is the observed width of a galaxy's spectral line (typically H\textsc{I}), serving as a proxy for its rotational velocity. Throughout, we refer to $\eta$ simply as the galaxy linewidth. We adopt the following quadratic form of the \ac{TFR}:
\begin{equation}\label{eq:TFR_absmag}
    M(\eta) =
    \begin{cases}
        \aTFR + \bTFR \eta + \cTFR \eta^2 &\mathrm{if}~\eta > 0\\
        \aTFR + \bTFR \eta &\mathrm{otherwise}
    \end{cases}
\end{equation}
where $\aTFR$ and $\bTFR$ are the zero-point and slope, respectively, and $\cTFR$ models the curvature of the relation for high-linewidth (i.e.~high-mass) galaxies.

\subsubsection{CosmicFlows-4 TFR sample}

We use the \ac{CF4} \ac{TFR} catalogue, a subset of the broader \ac{CF4} compilation~\citep{Tully_2023}, consisting of \num{9792} galaxies with $\zCMB \lesssim 0.05$ and no strict apparent magnitude limit~\citep{Kourkchi_2020B, Kourkchi_2020A}. Our analysis uses photometry exclusively in the all-sky \ac{WISE} W1 band (same as~\citetalias{Boubel_2025}), with additional selection criteria applied. We require $\eta > -0.3$ (to eliminate dwarf and low-mass galaxies, which may follow a different \ac{TFR} or have a higher scatter), Galactic latitude $|b| > 7.5^\circ$ to exclude the Galactic Zone of Avoidance, and a photometric quality flag of $5$ (``best''). After these cuts, the final W1 sample contains \num{3246} galaxies. The highest photometric quality requirement excludes $\sim$\num{1300} galaxies from the W1 subsample, while the Galactic latitude cut removes only 44 galaxies from the final sample. The Zone of Avoidance is typically defined as $|b| < 5^\circ$ (e.g.~\citealt{Staveley-Smith_1998}), making our choice relatively conservative. In contrast,~\citetalias{Boubel_2025} do not apply either cut. Since the publicly released \ac{CF4} catalogue does not provide magnitude uncertainties, we adopt a conservative fiducial uncertainty of $0.05~\mathrm{mag}$, following~\citet{Kourkchi_2019}. However, this uncertainty is subdominant to the intrinsic scatter in the \ac{TFR}, which is approximately $0.35~\mathrm{mag}$ (see~\cref{fig:C15_posterior_CF4_W1_mag_dipole} or~\cref{fig:C15_posterior_CF4_W1}). Moreover, unlike~\citetalias{Boubel_2025}, who impose a lower redshift cut of $c\zCMB > 3000~\kmsec$, we do not apply a lower redshift limit in our main analysis, but discuss its impact in~\cref{sec:discussion_conclusion}.

\subsubsection{2MTF sample}

Next, we consider the 2MASS Tully--Fisher Survey (2MTF), an all-sky sample of 2,062 spiral galaxies with \ac{TFR} measurements extending to redshifts of $z_{\mathrm{CMB}} \lesssim 0.03$~\citep{Masters_2008, Hong_2019}. The survey is selected in the $K$ band with an apparent magnitude limit of $K < 11.25$. We use the version of the catalogue compiled by~\citet{Boruah_2019}, which removes duplicates from the \SFI\ sample, includes only $K$-band photometry, and applies a quality cut on linewidths, retaining only galaxies with $-0.1 < \eta < 0.2$. These cuts leave \num{1247} galaxies.

\subsubsection{SFI\texttt{++} TFR sample}

We also consider the \SFI\ catalogue, an all-sky sample comprising galaxies and groups with TFR measurements extending to redshifts of $z_{\mathrm{CMB}} \lesssim 0.05$~\citep{Masters_2006, Springob_2007}. Unlike 2MTF, the survey does not impose a strict apparent magnitude limit, and photometry is provided in the $I$ band. In this work, we use the galaxy-only version of the catalogue compiled by~\citet{Boruah_2019}, who apply quality cuts to select galaxies within the~\TWOMPP\ footprint, impose a lower linewidth threshold to exclude low-mass galaxies, and use an iterative procedure to reject \ac{TFR} outliers. While~\citet{Boruah_2019} also impose a strict upper linewidth selection to ensure linearity of the TFR, we relax this constraint and allow for \ac{TFR} curvature, resulting in a final sample of \num{2010} galaxies.

\subsection{Pantheon\texttt{+} supernovae sample}

Type Ia \acp{SN} are widely used as standardisable candles in cosmology. The SALT2 model standardises their light curve~\citep{SALT2}, yielding a standardised apparent magnitude via the Tripp formula~\citep{Tripp_1998}:
\begin{equation}\label{eq:tripp_formula}
    m_{\rm standard} = m_{\rm obs} + \mathcal{A} x_1 - \mathcal{B} c,
\end{equation}
where $m_{\rm obs}$ is the observed \ac{SN} apparent magnitude, $x_1$ characterises the light curve stretch, and $c$ the colour. The global parameters $\mathcal{A}$ and $\mathcal{B}$ quantify the standardisation with respect to stretch and colour, respectively. Combined with the absolute magnitude $\MSN$, the standardised magnitude $m_{\rm standard}$ defines the distance modulus.

The \PP\ data set is a compilation of \num{1701} spectroscopically confirmed Type Ia \acp{SN} spanning redshifts from $z \sim 0.001$ to $\sim 2.3$~\citep{Scolnic_2022, Brout_2022, Peterson_2022, Carr_2022}. For consistency with our peculiar velocity modelling and to match the redshift range of the \ac{CF4} sample, we restrict the data to $\zCMB \leq 0.05$, yielding a subset of \num{525} \acp{SN}. \PP\ combines the original Pantheon sample~\citep{Scolnic_2018} with updated low- and high-redshift \acp{SN}, incorporating improved photometric calibration, light-curve standardisation, and systematic uncertainty modelling. Distance moduli are derived using the SALT2 light-curve fitter and corrected for selection effects via the BEAMS with Bias Corrections (BBC) method~\citep{Kessler_2017}, which introduces an additive bias correction term in~\cref{eq:tripp_formula}. We use such bias-corrected apparent magnitudes $m_{\rm corr}$ from the \PP\ sample, which include a fiducial Tripp calibration. We therefore sample only $\MSN$, not the light curve stretch and colour coefficients of the Tripp calibration.

Uncertainties in the distance moduli (or equivalently, bias-corrected apparent magnitude), including both statistical and systematic contributions, are encoded in a covariance matrix that incorporates uncertainty in the Tripp parameters $\mathcal{A}$ and $\mathcal{B}$, held fixed at fiducial values. The \PP\ data release therefore provides a set of standardised, bias-corrected apparent magnitudes and an associated covariance matrix, with the only global parameter varied in our analysis being $\MSN$, which is inferred jointly with the flow model.

\subsubsection{Lane et al. reanalysis}

A reanalysis of the \PP\ data was presented by~\cite{Lane_2024}, who constructed a \ac{SN} covariance matrix designed to minimise dependence on the assumed cosmology and peculiar velocities. Unlike the standard \PP\ covariance matrix and bias-corrected apparent magnitudes, which assume a fiducial cosmology and Tripp parameters, the~\citeauthor{Lane_2024} covariance matrix does not include any contribution from the Tripp parametrisation. This enables simultaneous inference of the Tripp coefficients, with the standardised apparent magnitudes expressed as in~\cref{eq:tripp_formula}, and no further bias corrections. We treat this as a variant of the \PP\ data, denoted ``\PPL'' henceforth.

\subsection{Peculiar velocity data}

In~\citetalias{Carrick_2015}, the luminosity-weighted density field is reconstructed from the redshift-space positions of galaxies in the \TWOMPP\ catalogue using the iterative method of~\citet{Yahil_1991}. \TWOMPP\ is a whole-sky redshift compilation of \num{69160} galaxies~\citep{Lavaux_2011}, derived from 2MASS photometry~\citep{Skrutskie_2006} and redshifts from 2MRS~\citep{Huchra_2012}, 6dF~\citep{Jones_2009}, and \ac{SDSS} DR7~\citep{Abazajian_2009}. Apparent magnitudes are corrected for Galactic extinction, $k$-corrections, evolution, and surface brightness dimming. The catalogue is magnitude-limited to $K < 11.5$ in the 2MRS region and $K < 12.5$ in the 6dF and \ac{SDSS} regions.

The velocity field is derived from the galaxy density field using linear theory and must be scaled to match peculiar velocities via a parameter $\beta^\star$, which we treat as a free parameter of the model. The field is generated on a $256^3$ grid with a box size of $400~\Mpch$, assuming $\Om = 0.3$. $\beta^\star$ is defined as
\begin{equation}\label{eq:beta_star}
    \beta^\star \equiv \frac{f \sigma_{8,\mathrm{NL}}}{\sigma_8^b},
\end{equation}
where $f$ is the dimensionless growth rate, with $f \approx \Omega_\mathrm{m}^{0.55}$ in $\Lambda$CDM~\citep{Bouchet_1995, Wang_1998}. The terms $\sigma_8^b$ and $\sigma_{8,\mathrm{NL}}$ represent the fluctuation amplitude in the biased galaxy field and in the non-linear matter field, respectively. The value of $\sigma_8^g$ in \TWOMPP\ was measured as $0.98 \pm 0.07$~\citep{westover} and $0.99 \pm 0.04$~\citep{Carrick_2015}.~\citetalias{Carrick_2015}, combined with peculiar velocity samples, has been extensively used to constrain the growth of structure and the $S_8$ parameter~\citep[e.g.][]{Boruah_2019, Said_2020, Boubel_2024,VF_olympics}.


\section{Methodology}\label{sec:method}

We adopt a forward modelling framework, predicting the observables from the model parameters and comparing them to the observed values through a likelihood function. This has the advantage that it exploits all the information in the data (as no summary statistics are used), and makes it straightforward to model systematics and other effects that impact the observed quantities. It also lends itself naturally to a Bayesian inference by determining the probability distributions of the parameters implied by the observational data.

In particular our method jointly calibrates the \ac{TFR} relation (or \acp{SN}) with the galaxy bias and the peculiar velocity field, following the methodology developed in~\citetalias{VF_olympics}. While \citetalias{Boubel_2025} also largely adopts a forward-modelling approach, one key step does not fall within that framework: by querying peculiar velocity in redshift space, they use the observed redshift to assign peculiar velocity to a galaxy, approximately updating the fiducial velocity field calibration of~\citetalias{Carrick_2015}, and then use that to estimate the cosmological redshift. That is a backward-modelling step (going from the redshift observation to a model parameter, the cosmological redshift or analogously distance) that introduces complications in triple-valued zones where a single observed redshift can map to multiple line-of-sight distances~\citep{StraussWillick_1995}. We instead model the observed redshift directly by querying the peculiar velocity field in real space. Here, we provide a brief summary of the model; a more detailed explanation is given in~\cref{sec:flow_model_full}, with further discussion in~\citetalias{VF_olympics}.

\subsection{Tully--Fisher model}\label{sec:flow_model_summary}

Each galaxy is described by its observed redshift, apparent magnitude, and linewidth, with which we infer the distance. The velocity field is modelled as a combination of a reconstructed internal flow, scaled by a parameter $\beta^\star$, and an external flow $\Vext$. In the absence of a velocity field model, the flow reduces to a constant $\Vext$ term. We also include a Gaussian dispersion parameter $\sigma_v$ to account for small-scale velocities not captured by the reconstruction.

The distances and true linewidths of each galaxy are latent model parameters over which we marginalise. The true linewidths are drawn from an inferred Gaussian distribution~\citep{MNR} and compared to the observations through a likelihood function. Via the \ac{TFR}, the true linewidths determine the absolute magnitudes, which, combined with the distances, yield predictions of the apparent magnitudes to be compared with the data. Likewise, the true model distances, together with the peculiar velocity model, provide predictions of the observed redshifts for comparison with the measurements. The forward model accounts for both homogeneous and inhomogeneous Malmquist bias. The homogeneous case is an assumption that sources are uniformly distributed in volume, yielding a distance prior $p(r) \propto r^2$ in the absence of selection effects. The inhomogeneous Malmquist bias is set by the density field of~\citetalias{Carrick_2015} and modelled as
\begin{equation}\label{eq:linear_bias}
    n(r) = 1 + b_1 \delta(r),
\end{equation}
where $\delta(r)$ is the density contrast at the galaxy's position. To ensure non-negative values, $n(r)$ is clipped from below at zero. We adopt this linear bias model to be consistent with the reconstruction of~\citetalias{Carrick_2015}, which uses linear theory to relate the galaxy density field, effectively smoothed over scales of $4~\Mpch$, to a peculiar velocity field. We treat $b_1$ as a model parameter and infer it with a wide uniform prior.

In general, \ac{TFR} samples are subject to a complex selection function combining H\textsc{I} flux, linewidth, optical magnitude, and possibly redshift selection. Following the empirical approach of~\cite{Lavaux_Virbius}, we model the prior distribution of galaxy distances as
\begin{equation}\label{eq:empirical_prior_distance}
    \pi(r \mid \bm{\theta})
    = \frac{n(r,\, \bm{\theta})\, f(r,\, \bm{\theta})}
           {\int \mathrm{d}r'\, n(r',\, \bm{\theta})\, f(r',\, \bm{\theta})},
\end{equation}
where $n(r,\, \bm{\theta})$ accounts for the inhomogeneous Malmquist bias through the large-scale density field, $\bm{\theta}$ collectively denotes the model parameters, and
\begin{equation}
    f(r,\, \bm{\theta}) = r^p \exp\!\left[-\left(\frac{r}{R}\right)^q\right].
\end{equation}
Here $p$, $q$, and $R$ are free parameters: $p \approx 2$ recovers the homogeneous Malmquist bias, $R$ sets the characteristic scale of sample incompleteness, and $q$ controls how sharply the completeness falls off. The normalisation in~\cref{eq:empirical_prior_distance} is computed explicitly since it depends on $\bm{\theta}$. This treatment remains phenomenological, approximating the selection rather than modelling it directly. A rigorous forward modelling of the survey selection, which requires knowledge of the selection function, is presented in~\cite{Kelly_2008} and we recently applied it to $H_0$ inference in~\cite{CH0}.

We extend the \ac{TFR} zero-point $\aTFR$ by introducing a dipole, such that
\begin{equation}\label{eq:TFR_dipole}
    \aTFR \rightarrow \aTFR + \bm{\Delta}_\mathrm{ZP} \cdot \hat{\bm{r}},
\end{equation}
where $\bm{\Delta}_\mathrm{ZP}$ is the zero-point dipole vector and $\hat{\bm{r}}$ is the unit vector in the direction of each galaxy. The dipole amplitude is denoted $\Delta_{\rm ZP}$, with direction specified in Galactic coordinates as $(\ell_{\Delta_{\rm ZP}},\,b_{\Delta_{\rm ZP}})$. The zero-point $\aTFR$ is perfectly degenerate with the Hubble constant $H_0$, since the distance modulus depends on the logarithm of the luminosity distance. This leads to a degenerate parameter combination $\aTFR + 5\log h$, where $h \equiv H_0 / (100~\kmsecMpc)$. Following~\citetalias{VF_olympics} and similar works, we express distances in units of $h^{-1}\,\mathrm{Mpc}$, rendering the analysis independent of $h$ and constraining only the degenerate combination.

As explored by~\citetalias{Boubel_2025}, one may allow for a dipole in this degenerate parameter across the sky. Provided sufficient sky coverage, such a dipole can be constrained. This introduces four possible interpretations:
\begin{enumerate}
    \item An anisotropic zero-point with isotropic $H_0$;
    \item An isotropic zero-point with an anisotropic $H_0$;
    \item A combination of both aforementioned effects;
    \item Spurious $H_0$ anisotropy arising from flows not captured by~\citetalias{Carrick_2015}.
\end{enumerate}
\citetalias{Boubel_2025} found no evidence for significant spatial variation in linewidths within the \ac{CF4} W1 subsample, by comparing the linewidth distribution of sources in the northern and southern hemispheres of the ALFALFA survey~\citep{Haynes_2018}. Moreover, the use of WISE photometry minimises systematic variation in magnitude calibration across the sky. Based on these considerations, they interpreted their measured dipole in $\aTFR + 5\log h$ as a potential signature of anisotropy in $H_0$. Assuming the dipole in $\aTFR + 5\log h$ arises from a dipole in $H_0$, the corresponding fractional variation in the Hubble constant is given by
\begin{equation}\label{eq:H0_dipole}
    \Delta H_0 / H_0 = 10^{\Delta_{\rm ZP} / 5} - 1,
\end{equation}
where $\Delta_{\rm ZP}$ is the magnitude of the dipole in the degenerate combination $\aTFR + 5\log h$ and $H_0$ is the monopole term of the Hubble constant.

As a second extension, we introduce a radial dependence of $\Vext$ to test for potential signatures of $H_0$ anisotropy. We uniformly sample the magnitude and sky direction of $\Vext$ at $N_{\rm knots}$ radial distance knots (set to $0,\,20,\,40,\,60,\,80$ and $100~\Mpch$ for \ac{CF4}), and apply cubic interpolation of its Cartesian components to evaluate $\Vext$ at each galaxy position. This procedure allows for a smoothly varying $\Vext$. Since a dipole in $H_0$ would imply a linearly increasing $\Vext$ with distance, this extension serves as a consistency check of the inferred $\bm{\Delta}_{\rm ZP}$ dipole. It also lets us infer the radial dependence of the bulk flow more generally.

\subsection{Pantheon\texttt{+} supernova model}\label{sec:flow_supernovae}

To model the \PP\ data, we adopt an approach analogous to that used for the \ac{TFR}, with the primary difference being that the predicted apparent magnitude is a function of the \ac{SN} properties. The \PP\ sample accounts for both statistical and systematic uncertainties, encapsulated in a covariance matrix for the apparent magnitudes. These uncertainties arise from the Tripp standardisation, photometric calibration, and the heterogeneity of the contributing surveys. While the full covariance matrix provided in the \PP\ release includes contributions from peculiar velocities, our model accounts for these explicitly. We therefore use a reduced version of the covariance matrix with the peculiar velocity terms removed, as provided to us by Anthony Carr (private communication).

As before, galaxy distances are sampled from the prior of~\cref{eq:empirical_prior_distance}. These are converted to distance moduli $\mu$, which combined with $\MSN$ give the predicted apparent magnitudes as $m_{\rm pred} = \mu + \MSN$ (the Tripp parameters are calibrated to fiducial values in the \PP\  sample). The likelihood is then evaluated against the standardised, bias-corrected magnitudes,
\begin{equation}
    \mathcal{L}(\bm{m}_{\rm corr} \mid \bm{m}_{\rm pred}) = \mathcal{N}(\bm{m}_{\rm corr};\bm{m}_{\rm pred},\, \mathbf{C}),
\end{equation}
where $\bm{m}_{\rm corr}$ is a vector of the standardised, bias-corrected apparent magnitudes, $\bm{m}_{\rm pred}$ is a vector of predicted apparent magnitudes, and $\mathbf{C}$ is the reduced \PP\ covariance matrix. After this, the rest of the inference follows the same steps as the \ac{TFR} analysis. Similarly as for the \ac{TFR}, we extend the standardised \ac{SN} absolute magnitude to include a dipole term:
\begin{equation}
    \MSN \rightarrow \MSN + \bm{\Delta}_{\rm ZP} \cdot \hat{\bm{r}}.
\end{equation}

On the other hand, when using the covariance matrix of \PP\ \acp{SN} from~\cite{Lane_2024}, we explicitly sample the Tripp parameters $\mathcal{A}$ and $\mathcal{B}$ to predict the \ac{SN} apparent magnitude (see Eq.~\ref{eq:tripp_formula}), since their data does not assume a fiducial Tripp standardisation or its contribution to the covariance matrix. We evaluate the likelihood in the apparent magnitudes as in Eq.~5 of~\cite{Seifert_2025}, except that we explicitly sample the galaxy distances, rather than setting them deterministically from the observed redshift which is equivalent to assuming no redshift uncertainty and no peculiar velocities.

\subsection{Galactic extinction}\label{sec:data_dustmap}

A potential systematic effect in inferring the dipole in the zero-point, or equivalently in the apparent magnitude, arises from the treatment of Galactic extinction. We test this on the~\ac{CF4} subsample, for which the applied Galactic extinction corrections are available in the public data release. The reported magnitudes in a given waveband $\lambda$ are given by (see Section~2.4 of~\citealt{Kourkchi_2019}):
\begin{equation}
    \overline{m}^\lambda = m_{\rm total}^\lambda - A_b^\lambda - A_k^\lambda - A_a^\lambda,
\end{equation}
where $m_{\rm total}^\lambda$ is the measured total magnitude, $A_b^\lambda$ is the Milky Way extinction, computed as
\begin{equation}
    A_b^\lambda = R_\lambda ~ \ebv,
\end{equation}
$A_k^\lambda$ is the $k$-correction, $A_a^\lambda$ is the total flux aperture correction, and $\ebv$ is the colour excess (reddening). As described by~\cite{Kourkchi_2019}, the \ac{CF4} catalogue uses the~\cite{Schlegel_1998} $100~\mu\mathrm{m}$ cirrus maps to extract the Milky Way $\ebv$ values. For wavebands $\lambda \in (u, g, r, i)$, the extinction coefficients $R_{\lambda}$ are taken from~\cite{Schlafly_2011}, while for the infrared bands, $R_{\rm W1} = 0.186$ and $R_{\rm W2} = 0.123$, as determined by~\cite{Fitzpatrick_1999}.

To assess potential systematics from Galactic extinction corrections, we use the \texttt{dustmaps} package\footnote{\url{https://dustmaps.readthedocs.io/en/latest/}}~\citep{dustmaps} to extract $\ebv$ values at the angular positions of \ac{CF4} galaxies. We adopt the extinction maps of~\citet{Chiang_2023} and~\citet{Planck_2016}. The former is a dust reddening map on the plane of the sky, derived from a reanalysis of~\citet{Schlafly_2011}, which in turn is based on~\citealt{Schlegel_1998}. It uses tomographically constrained templates from \ac{WISE} galaxy density fields to remove contamination from the cosmic infrared background (CIB). The latter map, from~\citet{Planck_2016}, employs the generalized needlet internal linear combination (GNILC) method to separate Galactic dust emission from CIB anisotropies, yielding an alternative all-sky extinction estimate. We consider these approaches for modelling Galactic extinction:
\begin{itemize}
    \item Use the $\ebv$ values from~\citet{Schlegel_1998}, jointly sampling the extinction coefficient $R_{\rm W1}$. We adopt a Gaussian prior on $R_{\rm W1}$ centred at 0.19, with either a standard deviation of 0.01, consistent with the measurement of~\citealt{Yuan_2013}, or a broader, more conservative prior with standard deviation 0.05.
    \item Use the $\ebv$ values from~\citet{Chiang_2023} or~\citet{Planck_2016}, also jointly sampling $R_{\rm W1}$ under the same prior choices as above.
\end{itemize}

The other \ac{TFR} samples, 2MTF and \SFI, are compiled in the optical and hence may be more susceptible to dust extinction. We cannot test this explicitly as the dust corrections they used are not publicly available, but note that the very small effect between different dust models found for CF4 suggests that the effect still would not be significant. Similarly, we do not consider Galactic extinction variations in the \PP\ samples.

\subsection{Mock data generation}\label{sec:mock_data_generation}

We use mock data to estimate the sample size required for a \ac{CF4}-like survey to yield a significant detection of a dipole of given strength. The mock catalogue is designed to replicate the \ac{CF4} \ac{TFR} subsample with W1 photometry and full-sky coverage, excluding the Galactic Zone of Avoidance. The injected parameter values, listed in~\cref{tab:mock_TFR_injected_values}, correspond to the posterior mean from the \ac{CF4} inference.

\begin{figure}
    \centering
    \includegraphics[width=\columnwidth]{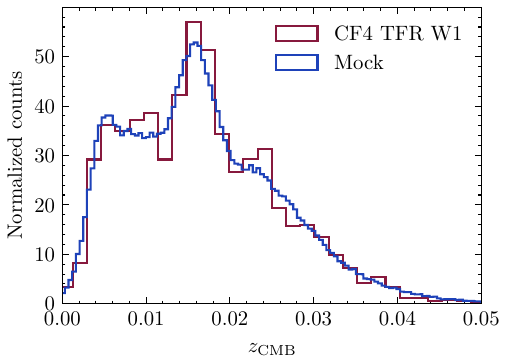}
    \caption{Comparison of the redshift distribution of the \ac{CF4} \ac{TFR} W1 sample with a mock catalogue generated using the parameters in~\cref{tab:mock_TFR_injected_values}. The mock is designed to replicate the \ac{CF4} subsample and assess how the dipole detection significance depends on sample size.}
    \label{fig:mock_redshift}
\end{figure}

For $N$ sources, we draw sky positions uniformly over the sphere, excluding a mock ``Zone of Avoidance'' defined by $|b| < 7.5^\circ$. Each line of sight is then used to query the~\citetalias{Carrick_2015} field for the density and peculiar velocity. The radial distance $r$ is sampled from the prior in~\cref{eq:empirical_prior_distance}, which incorporates the linear galaxy bias of~\cref{eq:linear_bias}. At this distance we evaluate the line-of-sight peculiar velocity $\bm{v}(\bm{r})$ and compute the true redshift,
\begin{equation}
    1 + \ztrue = \big(1 + \zcosmo(r)\big)
    \left(1 + \frac{\left[\Vext + \beta \bm{v}(\bm{r})\right] \cdot \hat{\bm{r}}}{c}\right),
\end{equation}
where $\zcosmo$ is the cosmological redshift at distance $r$ and $\hat{\bm{r}}$ is the unit vector toward the source. The observed redshift is then drawn with scatter $\sigma_v$,
\begin{equation}
    c z_{\rm obs} \hookleftarrow \mathcal{N}(c \ztrue,\,\sigma_v),
\end{equation}
assuming the ``measurement'' error of $z_{\rm obs}$ to be subdominant to $\sigma_v$. The apparent magnitude is obtained by first sampling $\etatrue$ from a Gaussian hyperprior,
\begin{equation}
    \etatrue \hookleftarrow \mathcal{N}(\hat{\eta},\, w_\eta),
\end{equation}
and then the observed linewidth as
\begin{equation}
    \etaobs \hookleftarrow \mathcal{N}(\etatrue,\, \sigma_\eta).
\end{equation}
The true apparent magnitude is
\begin{equation}
    \mtrue = \mu(r) + M(\etatrue),
\end{equation}
where $\mu(r)$ is the distance modulus and $M(\etatrue)$ is defined in~\cref{eq:TFR_absmag}. The observed magnitude is drawn as
\begin{equation}
    \mobs \hookleftarrow \mathcal{N}\left(\mtrue,\, \sqrt{\sigma_m^2 + \sint^2}\right).
\end{equation}
The resulting catalogue contains observed redshifts, magnitudes, linewidths, and sky positions. An example redshift distribution compared with the \ac{CF4} data is shown in~\cref{fig:mock_redshift}.

\subsection{Inference procedure}\label{sec:inference_procedure}

To sample the posterior distribution we use the \acl{NUTS} (\ac{NUTS};~\citealt{Hoffman_2011}) method of \ac{HMC} as implemented in the \texttt{numpyro}\footnote{\url{https://num.pyro.ai/en/latest/}} package~\citep{Phan_2019, Bingham_2019}, ensuring a Gelman--Rubin statistic $\hat{R}-1 \leq 0.001$ for convergence~\citep{Gelman_1992}.
For each model, we run four independent chains with \num{1000} warm-up steps and \num{2000} sampling steps, typically yielding more than \num{4000} effective samples for the dipole and other parameters.
We adopt a uniform prior on the zero-point dipole amplitude from 0 to 0.3 and sample its direction uniformly over the surface of a sphere. For all remaining parameters, we adopt uniform priors over sufficiently large ranges, except for $\sigma_v$, $\sint$ for which we adopt reference (and scale invariant) priors $\pi(x) \propto 1/x$. All these parameters are shared between both the isotropic and dipole models so their choice does not affect the Bayes factors between the models.

We calculate the Bayesian evidence to quantify models' relative goodness-of-fit. This is the probability of the data $D$ given the model $M$:
\begin{equation}
    \mathcal{Z}\equiv p(D \mid M) = \int \dd\bm{\theta} \: \mathcal{L}(D\mid\bm\theta,\,M) \: \pi(\bm\theta)
\end{equation}
where $\bm\theta$ is the set of free parameters of the model, $\mathcal{L}$ is the likelihood function and $\pi(\bm\theta)$ is the prior on the parameters. We estimate this using the \texttt{harmonic} package\footnote{\url{https://github.com/astro-informatics/harmonic}}, which fits a normalising flow to the posterior samples to compute evidence using the harmonic estimator~\citep{McEwan_2021,Polanska_2024}. For the \ac{TFR} samples we numerically marginalise over both galaxy latent parameters ($r$ and $\etatrue$) at each \ac{MCMC} step, yielding a low-dimensional posterior suitable for evidence computation with \texttt{harmonic}. In contrast, for the \PP\ samples the covariance prevents straightforward numerical marginalisation of the distances, requiring them to be sampled explicitly. This results in an approximately $600$-dimensional posterior, which makes evidence calculation challenging.
While \texttt{harmonic} is designed to compute the Bayesian evidence directly from posterior samples of moderately high dimension, it has typically been tested on $\mathcal{O}(10)$-dimensional problems. For example,~\cite{Piras_2024} applied \texttt{harmonic} to a 39-dimensional posterior, validating their results against nested sampling, and also to a 159-dimensional posterior for which nested sampling was computationally infeasible. \texttt{harmonic} uses a normalising flow within a harmonic-mean framework to perform density estimation for evidence computation, but for a $\sim600$-dimensional posterior the required number of samples becomes prohibitive, convergence cannot be guaranteed, and its use has not been previously explored in such high dimensionality. Accordingly, we do not apply \texttt{harmonic} in this high-dimensional setting.
We therefore restrict evidence computation to the \ac{TFR} samples. For these, we compare models using the Bayes factor, defined as the ratio of evidences for two models $M_1$ and $M_2$,
\begin{equation}
    \mathcal{B} \equiv \frac{p(D\mid M_1)}{p(D\mid M_2)},
\end{equation}
which quantifies the relative probability of the data under the two models. Assuming equal prior probabilities for the isotropic and dipole models, the Bayes factor is equal to the model odds $p(M_1 \mid D) / p(M_2 \mid D)$.

\begin{table}
    \centering
    \begin{tabularx}{\columnwidth}{llp{4cm}}
    \hline\hline
    Parameter                   & Injected Value                          & Description \\
    \hline
    $a_{\rm TFR}$               & $-19.95$                                & TFR zero-point \\
    $b_{\rm TFR}$               & $-9.6$                                  & TFR slope \\
    $c_{\rm TFR}$               & $10.5$                                  & TFR curvature for $\eta > 0$ \\
    $\sint$                     & $0.34$                                  & Intrinsic TFR scatter \\
    $\sigma_\eta$               & $0.02$                                  & Linewidth uncertainty \\
    $\sigma_m$                  & $0.05$                                  & Magnitude uncertainty \\
    $b_{\min}$                  & $7.5^\circ$                             & Galactic latitude cut: $|b| > b_{\min}$ \\
    $V_{\mathrm{ext}}$          & $160~\kmsec$                             & External flow magnitude \\
    $\ell_{\mathrm{ext}}$       & $302^\circ$                             & External flow Galactic longitude \\
    $b_{\mathrm{ext}}$          & $-17^\circ$                             & External flow Galactic latitude \\
    $\Delta_{\rm ZP}$           & $0.087$                                 & Zero-point dipole amplitude \\
    $\ell_{\Delta_{\rm ZP}}$ & $127^\circ$                             & Zero-point dipole Galactic longitude \\
    $b_{\Delta_{\rm ZP}}$   & $10^\circ$                               & Zero-point dipole Galactic latitude \\
    $\sigma_v$                  & $280~\kmsec$               & Redshift uncertainty including small-scale velocity dispersion \\
    $\beta^\star$                     & $0.43$                                  & Velocity scaling of~\protect\citetalias{Carrick_2015} \\
    $p$                         & $1.8$                                   & Distance prior low-distance exponent\\
    $R$                         & $32~\Mpch$                               & Distance prior incompleteness distance\\
    $q$                         & $1.4$                                   & Distance prior incompleteness rate \\
    \hline\hline
    \end{tabularx}
    \caption{Injected parameter values used to generate mock \ac{TFR} catalogues mimicking the \ac{CF4} W1-band sample, assuming the \citetalias{Carrick_2015} density and peculiar velocity field. The parameters correspond to the posterior mean from the dipole inference on the observed sample. A comparison of the resulting redshift distribution for one mock catalogue with the \ac{CF4} subsample is shown in~\cref{fig:mock_redshift}.}
    \label{tab:mock_TFR_injected_values}
\end{table}

\section{Results}\label{sec:results}

We present results from applying our flow model to various data sets, focusing primarily on the \ac{CF4} \ac{TFR} W1 subsample. In~\cref{sec:main_results}, we report constraints on the zero-point dipole magnitude and direction under different Galactic dust corrections and extinction priors, and assess the model preference based on Bayesian evidence. In~\cref{sec:results_2mtf_sfi}, we apply the same model to the 2MTF and \SFI\ \ac{TFR} samples. We then examine the \PP\ \ac{SN} sample in~\cref{sec:results_pp}.
In~\cref{tab:dipole_summary} we summarise the recovered dipole, external flow, and Bayes factors for the \ac{CF4} W1 band, 2MTF, \SFI, and the \PP\ samples, while in~\cref{fig:dipole_comparison} we compare the inferred dipole direction with literature estimates, finding broad consistency. In~\cref{sec:variations} we investigate the robustness of these results to assumptions about peculiar velocity corrections and the prior on the dipole amplitude. In~\cref{sec:mock_results}, we use mock \ac{TFR} catalogues to assess the detectability of a zero-point dipole as a function of survey sample size, evaluating the level of anisotropy that could be robustly inferred given realistic distance uncertainties. Finally, in~\cref{sec:implied_bulk_flow} we examine the bulk flow equivalent to the $H_0$ dipole, and compare it with both the bulk flow inferred when $\Vext$ is allowed a radial dependence and with \ac{LCDM} expectation.

\subsection{CosmicFlows-4 TFR W1 dipole}\label{sec:main_results}

\begin{figure*}
    \centering
    \includegraphics[width=\textwidth]{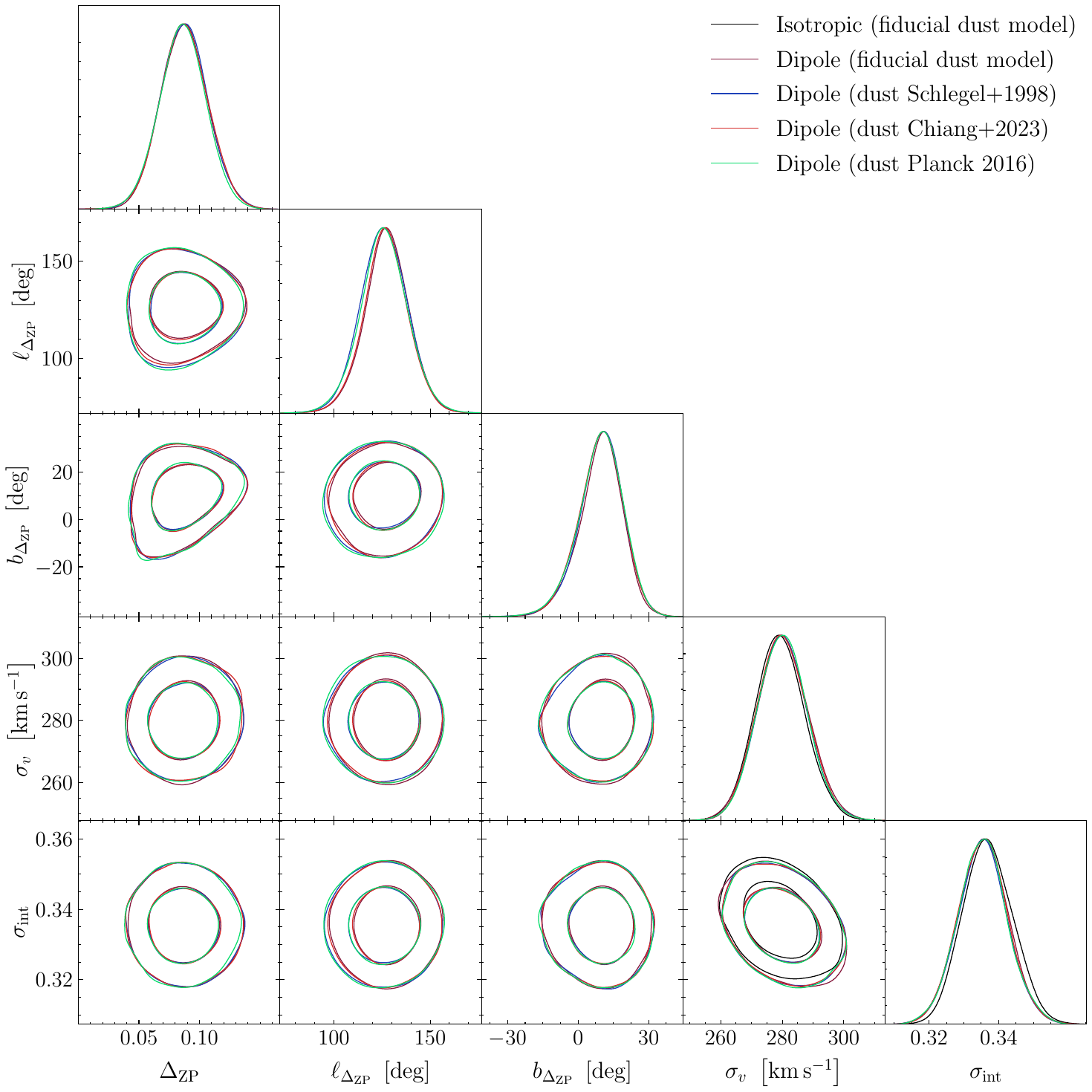}
    \caption{Comparison of the inferred \ac{TFR} zero-point dipole under different Galactic dust corrections using the \ac{CF4} \ac{TFR} W1 data: (a) fixed extinction from the \ac{CF4} catalogue which is based on ~\protect\cite{Schlegel_1998}, (b) $\ebv$ from~\protect\cite{Schlegel_1998} with a sampled extinction coefficient $R_{\rm W1}$, (c) $\ebv$ from the~\protect\cite{Chiang_2023} map, and (d) $\ebv$ from the~\protect\cite{Planck_2016} map, both also with sampled $R_{\rm W1}$. The recovered zero-point dipole is unaffected by the extinction treatment and its magnitude is $\Delta_{\rm ZP} = 0.087 \pm 0.19$ mag. The values of redshift scatter $\sigma_v$ and \ac{TFR} scatter $\sint$ show no improvement over the isotropic model without a zero-point dipole, though the Bayesian evidence favours the anisotropic models with odds of $877\!:\!1$.}
    \label{fig:C15_posterior_CF4_W1_mag_dipole}
\end{figure*}

Applying our model to the \ac{CF4} \ac{TFR} W1 sample, we infer a zero-point dipole with magnitude of $0.087 \pm 0.019$ mag and direction of $(\ell, b) = (127 \pm 11^\circ, 10\pm 8^\circ)$ in Galactic coordinates, corresponding to a dipole in $H_0$ of $4.1\pm0.9$ per cent. We now focus on the marginal posterior for selected parameters; the full posterior is discussed in~\cref{sec:full_posterior}. \Cref{fig:C15_posterior_CF4_W1_mag_dipole} shows the recovered posteriors for the zero-point dipole amplitude and direction, as well as for $\sigma_v$ and $\sint$, which represent the velocity dispersion and intrinsic \ac{TFR} scatter, respectively. Both $\sigma_v$ and $\sint$ quantify the residuals of the model: a model that describes the data better is likely to yield smaller values of $\sigma_v$ and $\sint$. We consider three cases:
\begin{enumerate}
    \item Isotropic \ac{TFR} zero-point.
    \item Dipole in the zero-point with fixed extinction from the \ac{CF4} catalogue.
    \item Dipole in the zero-point with the extinction coefficient jointly inferred from the data, using one of three dust maps:~\cite{Schlegel_1998},~\cite{Chiang_2023}, or~\cite{Planck_2016}.
\end{enumerate}

We find that neither the choice of dust map nor the sampling of the Galactic extinction coefficient has an appreciable effect on the inferred zero-point dipole. This indicates no significant degeneracy between the dipole and the Galactic extinction correction. We also consider two prior choices for the extinction coefficient $R_{\rm W1}$ (a narrow prior consistent with~\citet{Yuan_2013}, and a broader, less informative alternative) and find that the results are similarly insensitive to this choice. All three models that include a dipole in the zero-point yield values of $\sigma_v$ and $\sint$ that are near-identical to those recovered by the isotropic model without a dipole. This would suggest that introducing a dipole does not appreciably improve the fit or capture additional trends in the data. However, upon considering a model comparison we find that the Bayesian evidence favours the anisotropic model with $\log \mathcal{B} = 2.938$, or odds of $877\!:\!1$. We also find that the evidences of the models with different extinction treatments are nearly identical, indicating that the \ac{CF4} \ac{TFR} W1 data is insufficient to distinguish between the dust maps.

It is important to note that, since the dipole magnitude is a strictly positive quantity, even if the Cartesian components of the dipole vector are consistent with zero, the magnitude of the dipole can still be significantly positive (for example, if the components follow a zero-mean Gaussian, the magnitude of the vector follows a Maxwell–Boltzmann distribution). The fact that the posterior of the dipole amplitude is a few $\sigma$ inconsistent with zero is therefore not a reliable indicator of the presence of a dipole, hence the necessity of computing the Bayes factor.

Our conclusions are consistent with those of~\citetalias{Boubel_2025}, who report a dipole magnitude of $0.063 \pm 0.016$ mag, slightly lower than our inferred value and with a smaller quoted uncertainty. They find only a very weak support for the anisotropic model, reporting $\log \mathcal{B} = 0.43$ in favour of the model including a zero-point dipole. However, they derive the Bayes factor using an approximation based on the Bayesian Information Criterion. We compare our results to~\citetalias{Boubel_2025} further in~\cref{sec:discussion_conclusion}.

\subsection{2MTF and SFI\texttt{++} dipole}\label{sec:results_2mtf_sfi}

We next consider the two alternative \ac{TFR} samples: 2MTF and \SFI. While subsets of these were used in constructing the \ac{CF4} \ac{TFR} sample, we now analyse them independently. We apply the same flow model as for the \ac{CF4} \ac{TFR} W1 subsample, using identical priors on the zero-point dipole parameters (uniform in magnitude and direction on the sphere). We show the results in~\cref{fig:C15_2MTF_SFI_posterior}. For 2MTF, we infer a dipole magnitude of $0.040 \pm 0.024$ mag with the evidence marginally favouring the isotropic model, with $\log \mathcal{B} = -0.381$. For \SFI, the recovered dipole magnitude is $0.051 \pm 0.018$ mag, though this time the evidence ratio marginally favours the inclusion of the dipole with $\log \mathcal{B} = 0.465$. Compared to \SFI, 2MTF covers a smaller redshift range and contains fewer galaxies, so it is unsurprising that it is less informative regarding the zero-point dipole.

\subsection{Pantheon\texttt{+} dipole}\label{sec:results_pp}

\begin{figure*}
    \centering
    \begin{subfigure}[t]{0.48\textwidth}
        \centering
        \includegraphics[width=\textwidth]{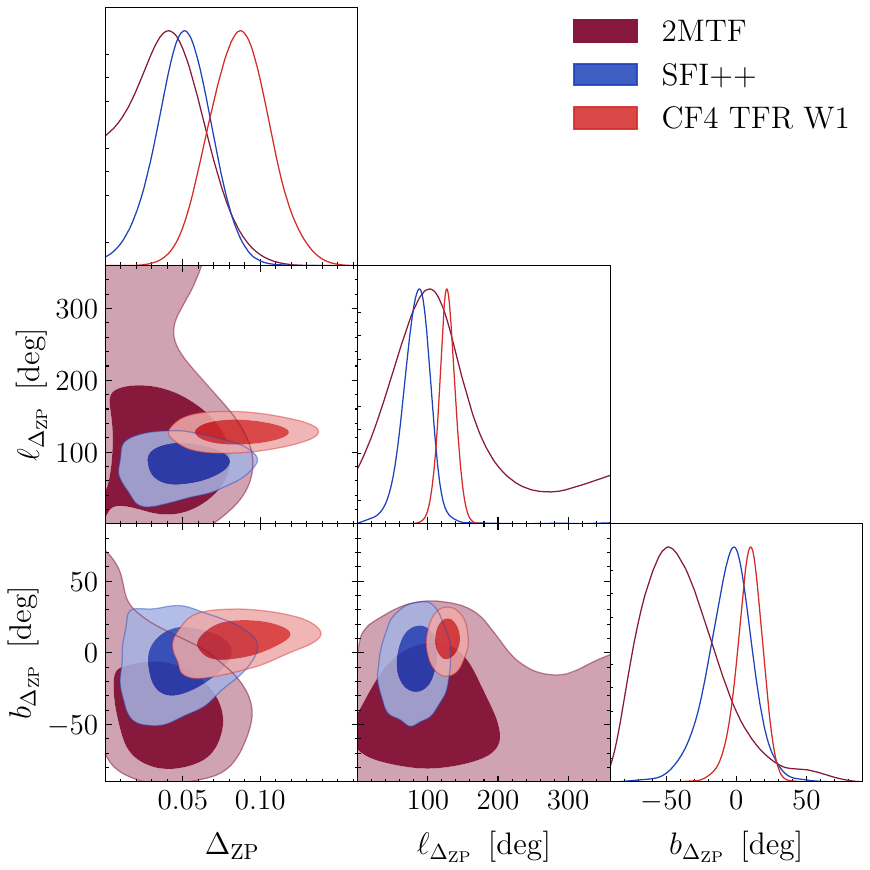}
        \caption{Inferred zero-point dipole in the \ac{TFR} samples. Except for 2MTF (the least informative sample), the dipole magnitude is significantly offset from zero.}
        \label{fig:C15_2MTF_SFI_posterior}
    \end{subfigure}\hfill
    \begin{subfigure}[t]{0.48\textwidth}
        \centering
        \includegraphics[width=\textwidth]{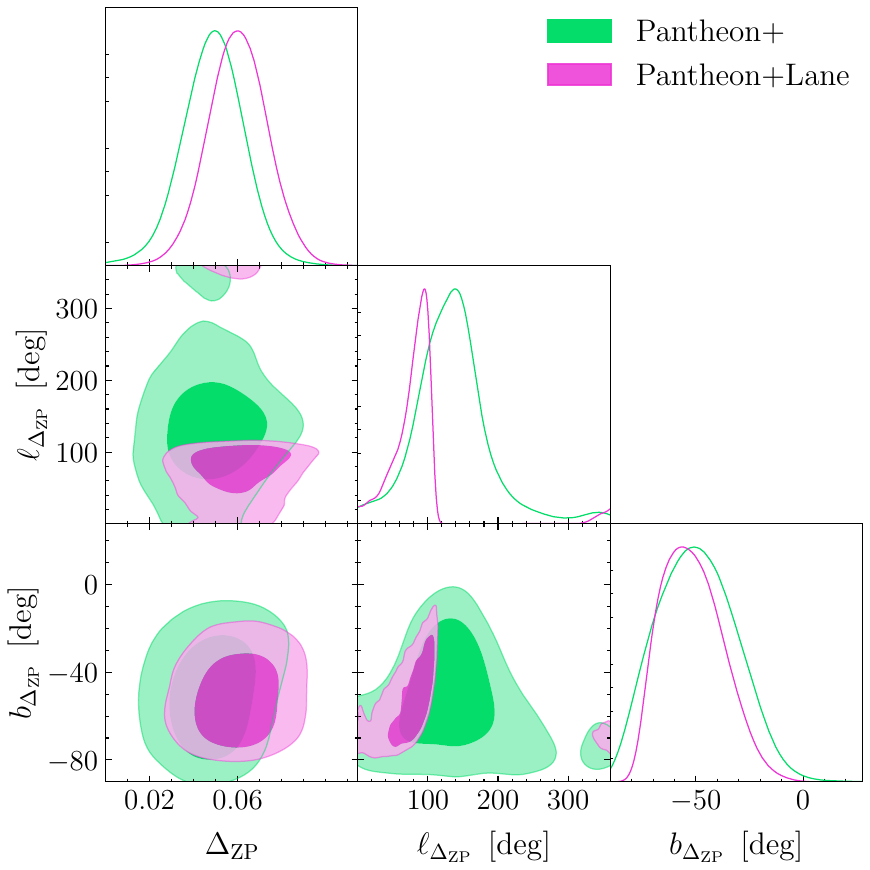}
        \caption{Inferred zero-point dipole in the \ac{SN} sample. The dipole directions from the \ac{TFR} and \ac{SN} samples differ only slightly in the $b_{\Delta_{\rm ZP}}$ coordinate.}
        \label{fig:C15_SN_posterior}
    \end{subfigure}
    \caption{Comparison of the inferred zero-point dipole in the \ac{TFR} and \ac{SN} samples. The contours show the $1\sigma$ and $2\sigma$ confidence regions.}
    \label{fig:C15_TFR_SN}
\end{figure*}

Now we test the presence of a dipole in the \PP\ data, repeating the procedure described above with the \ac{SN} flow model described in~\cref{sec:flow_supernovae}. Owing to correlated uncertainties in \ac{SN} standardisation, the sampled distances cannot be (easily) numerically marginalized, resulting in a high-dimensional posterior and we do not compute the Bayesian evidence. In~\cref{fig:C15_SN_posterior} we show that for the \PP\ sample the recovered dipole amplitude is $0.049 \pm 0.013$ mag, which is nearly $4\sigma$ inconsistent with zero. In the \PPL\ sample the dipole amplitude is only marginally higher: $0.060 \pm 0.13$ mag. Since the dipole magnitudes are significantly different from zero, the data appear to favour the inclusion of a dipole even though we cannot compute the Bayes factor. The inferred dipole directions of the two samples are mutually consistent but show some tension with the \ac{TFR} samples, possibly due to their different redshift depths.

Furthermore, we find that the intrinsic \ac{SN} apparent magnitude scatter $\sint$ inferred by the model is $0.01 \pm 0.01$ for \PP\ and $0.14 \pm 0.01$ mag for \PPL (both with and without the zero-point dipole). \cite{Brout_2022} propagate an additional intrinsic scatter term directly into the covariance matrix, which explains why no further scatter is required (indeed the propagated scatter may be overestimated). \cite{Sah} argue that the \PP\ sample may suppress deviations from \ac{LCDM} through this treatment, contrasting it with the covariance matrices of the Joint Light-Curve Analysis (JLA;~\citealt{Betoule_2014}) and \PPL~\citep{Lane_2024}, which have a much smaller diagonal scatter. However, our inference of an additional intrinsic scatter of $0.14 \pm 0.01$ mag indicates that the extra scatter propagated into the \PP\ covariance is justified.

\begin{table*}
\centering
\begin{tabular}{lcccccccc}
\hline
Sample &
$\Delta_\mathrm{ZP}$ &
$\ell_{\Delta_\mathrm{ZP}}~[\mathrm{deg}]$ &
$b_{\Delta_\mathrm{ZP}}~[\mathrm{deg}]$ &
$V_\mathrm{ext}~[\mathrm{km}\,\mathrm{s}^{-1}]$ &
$\ell_\mathrm{ext}~[\mathrm{deg}]$ &
$b_\mathrm{ext}~[\mathrm{deg}]$ &
$\Delta H_0 / H_0$ &
$\log \mathcal{B}$ \\
\hline
CF4 W1        & $0.087 \pm 0.019$ & $128 \pm 12$ & $9 \pm 10$   & $158 \pm 20$ & $303 \pm 8$ & $-17 \pm 5$ & $0.041 \pm 0.009$ & $+2.94$ \\
SFI++         & $0.051 \pm 0.018$ & $84 \pm 25$  & $-5 \pm 17$  & $168 \pm 23$ & $311 \pm 7$ & $4 \pm 6$   & $0.024 \pm 0.008$ & $+0.47$ \\
2MTF          & $0.041 \pm 0.023$ & $130 \pm 83$ & $-37 \pm 29$ & $150 \pm 21$ & $317 \pm 9$ & $0 \pm 7$   & $0.019 \pm 0.011$ & $-0.38$ \\
Pantheon+     & $0.049 \pm 0.014$ & $135 \pm 56$ & $-49 \pm 18$ & $212 \pm 39$ & $312 \pm 10$& $-9 \pm 7$  & $0.023 \pm 0.006$ & -- \\
Pantheon+Lane & $0.060 \pm 0.014$ & $84 \pm 50$  & $-52 \pm 14$ & $220 \pm 34$ & $312 \pm 8$ & $-21 \pm 6$ & $0.028 \pm 0.006$ & -- \\
\hline
\end{tabular}
\label{tab:dipole_summary}
\caption{Summary of zero-point dipole and external flow parameters, as well as the fractional anisotropy in $H_0$, and Bayes factors in favour of the anisotropic model. Positive values of $\log \mathcal{B}$ indicate preference for the anisotropic model, negative values for the isotropic model. Only values with $|\log \mathcal{B}| \geq 2$ are considered as ``decisive'' evidence~\protect\citep{Jeffreys}.}
\end{table*}

\begin{figure*}
    \centering
    \includegraphics[width=\textwidth]{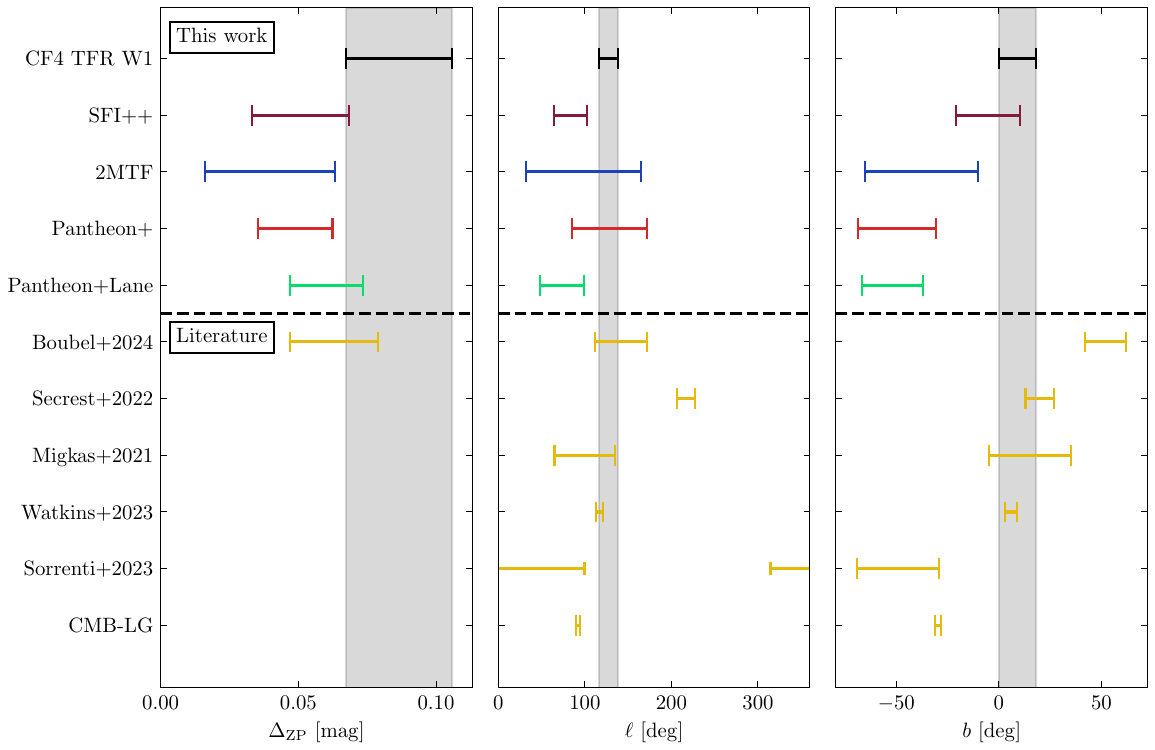}
    \caption{Summary of the inferred zero-point dipole for each catalogue, given by its magnitude $\Delta_{\rm ZP}$ and direction in Galactic coordinates $(\ell, b)$, compared with literature estimates: the \ac{CF4} \ac{TFR} measurement~\protect\citep{Boubel_2025}, the quasar dipole~\protect\citep{Secrest_2022}, the cluster scaling-relation dipole~\protect\citep{Migkas_2021}, the \ac{CF4} bulk flow~\protect\citep{Watkins_2023}, the \PP\ dipole (\protect\citealt{Sorrenti_2023}, their $z < 0.05$ sample), and the Local Group velocity in the \ac{CMB} frame~\protect\citep{Planck_2020}. For the latter four, we plot the opposite direction to that reported, as our dipole is defined in the zero-point rather than in peculiar velocities. We show the inferred magnitude only for~\protect\cite{Boubel_2025}, as the remaining works measure dipoles in different quantities that we do not convert. All results are reported in the \ac{CMB} frame. Error bars indicate the 16\textsuperscript{th} and 84\textsuperscript{th} percentiles.}
    \label{fig:dipole_comparison}
\end{figure*}

\subsection{Model variations}\label{sec:variations}

Here we briefly consider how the results change on altering two aspects of the model. First, we consider the case in which the~\citetalias{Carrick_2015} reconstruction is not used to model the underlying density field and peculiar velocities. In this scenario, we omit the inhomogeneous Malmquist bias and model the velocity field with a constant flow $\Vext$. We then repeat the inference using the \ac{CF4} \ac{TFR} W1 subsample. In this case, the recovered dipole amplitude is $0.179 \pm 0.023$ mag, significantly larger than the value obtained when accounting for peculiar velocities. The model selection in this ``no peculiar velocity'' scenario favours the inclusion of a zero-point dipole, effectively compensating for the unmodelled peculiar velocities, with a Bayes factor of $\log \mathcal{B} = 6.73$. On the other hand, for \PP\ we only infer a significantly larger zero-point dipole uncertainty when not accounting for peculiar velocities, with the dipole magnitude being $0.051 \pm 0.035$ mag. This highlights the importance of robustly accounting for peculiar velocities when investigating anisotropy, although it is important to bear in mind that these corrections may also absorb some anisotropy that may be present (as we discuss in~\cref{sec:intro}).

\begin{figure}
    \centering
    \includegraphics[width=\columnwidth]{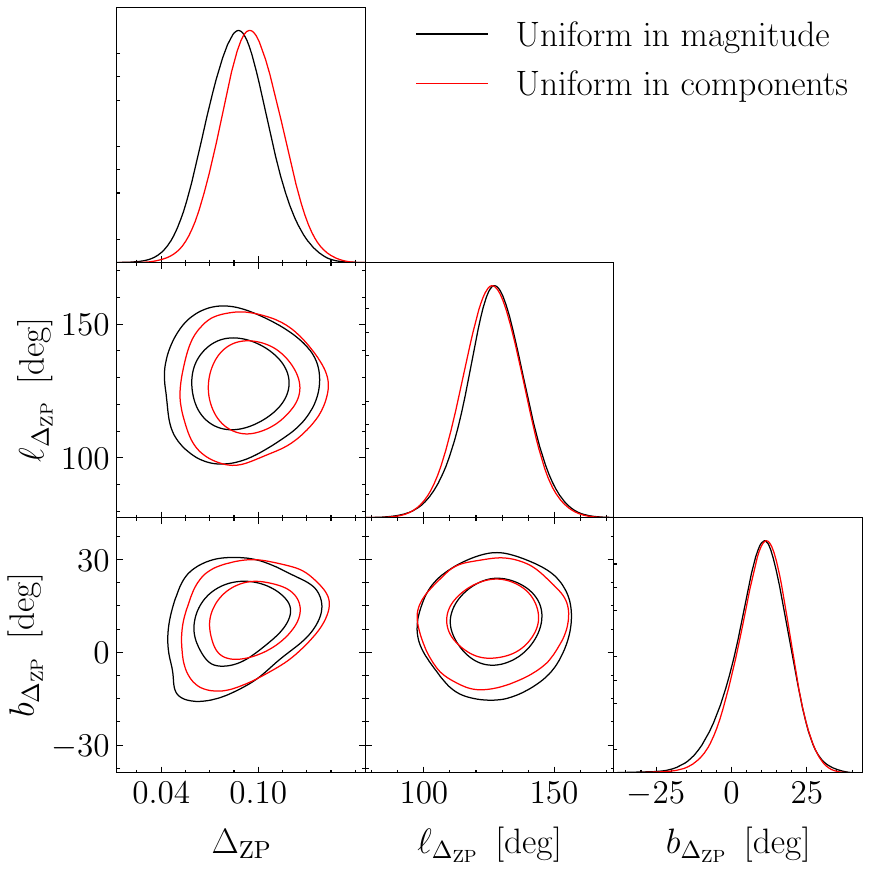}
    \caption{Comparison of the inferred dipole in the \ac{CF4} \ac{TFR} W1 subsample under two different priors: uniform in the dipole magnitude or uniform in Cartesian components of the dipole. The latter induces a prior that favours larger magnitudes, resulting in a higher posterior dipole amplitude. The inferred direction is largely insensitive to the prior choice. Contours show the $1\sigma$ and $2\sigma$ confidence regions.}
    \label{fig:prior_dependence}
\end{figure}

Second, we consider the impact of the assumed prior on the dipole amplitude. Previously we have taken this to be a uniform prior on the magnitude of the dipole vector; we now make this uniform on the Cartesian components (the choice made by~\citetalias{Boubel_2025}). This implicitly induces a prior on the dipole magnitude proportional to the square of the magnitude, pushing the posterior towards larger amplitudes. We repeat the analysis on the \ac{CF4} \ac{TFR} W1 subsample using this component-based prior, and compare the results to those obtained with a uniform prior on the magnitude in~\cref{fig:prior_dependence}. As expected, the recovered dipole magnitude is larger under the component prior, while the inferred direction is essentially unchanged. We find that the dipole magnitude is $0.095 \pm 0.019$ mag with the flat-component prior, compared to $0.087 \pm 0.019$ mag with the flat-magnitude prior, a $\sim0.5\sigma$ difference.

\subsection{Dipole mock analysis}\label{sec:mock_results}

Above we reported a significant preference for a dipole in the \ac{CF4} \ac{TFR} W1 sample. We now inject a dipole of magnitude $0.087\,\mathrm{mag}$, corresponding to the posterior mean value inferred from the \ac{CF4} sample, and examine how the evidence ratio between anisotropic and isotropic models depends on sample size. The remaining injected parameters are listed in~\cref{tab:mock_TFR_injected_values}, and an example redshift distribution of a mock sample is shown in~\cref{fig:mock_redshift}. Following~\cref{sec:mock_data_generation}, we generate mock catalogues with \num{500}, \num{1000}, \num{2000}, \num{4000}, \num{8000}, \num{16000}, and \num{32000} sources, drawing ten random realisations for each size. We then apply our flow model to the mock \ac{TFR} data, fitting both an isotropic model and one including a zero-point dipole, and compute the corresponding evidence ratio.

\begin{figure}
    \centering
    \includegraphics[width=\columnwidth]{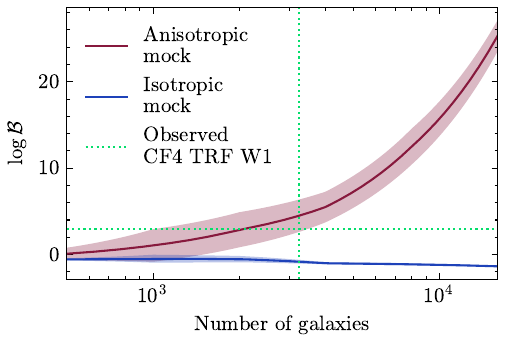}
    \caption{Model comparison using \ac{CF4} \ac{TFR} W1-like mock data sets of varying size (see Section~\ref{sec:mock_data_generation}), testing an anisotropic zero-point dipole model against an isotropic alternative. Mocks are generated either with a zero-point dipole of magnitude $0.087$ (equal to the posterior mean of our constraints; red) or without it (blue). Larger values indicate stronger preference for the anisotropic model. The green lines indicate the sample size and Bayes factor of the \ac{CF4} \ac{TFR} W1 data, which are sufficient to yield ``decisive'' evidence on the Jeffreys' scale ($\log \mathcal{B} > 2$,~\protect\citealt{Jeffreys}). The shaded band shows the 16\textsuperscript{th}–84\textsuperscript{th} percentile range of the Bayes factor $\log \mathcal{B}$ across mock realisations of the same size.}
    \label{fig:mock_evidences}
\end{figure}

We present these results in~\cref{fig:mock_evidences}. The Bayes factor becomes decisive ($\log \mathcal{B} > 2$ on the Jeffreys' scale,~\citealt{Jeffreys}) in favour of the dipole model only at sample sizes of order \num{2000}. At smaller sizes it remains inconclusive and does not reliably discriminate between the models. Since the W1 sample after our quality cuts contains \num{3246} galaxies, it is reassuring that the Bayes factor favours the zero-point dipole model near-decisively at that scale, assuming the injected dipole magnitude. The expected Bayes factor preference of mock samples of similar size to \ac{CF4} is likewise consistent with our results on the data. \cref{fig:mock_evidences} is also consistent with the absence of a significant dipole detection in the smaller 2MTF and \SFI\ samples. For mock samples without an injected dipole the Bayes factor is always negative, favouring the isotropic model, although the strength of this preference increases only weakly with sample size. We note that these results are specific to a \ac{CF4} \ac{TFR} W1-like survey: while deeper surveys could in principle provide stronger constraints through larger samples, potential gains may be offset by increased absolute distance uncertainties at higher redshift.

\subsection{Bulk flow measurement}\label{sec:implied_bulk_flow}

The inferred dipole in $\bm{\Delta}_{\rm ZP}$ can be interpreted as a dipole in $H_0$ (see Eq.~\ref{eq:H0_dipole}), or alternatively as a bulk flow linearly rising with distance under a cosmological model with isotropic expansion, which in turn corresponds to a bulk flow that increases linearly with distance. The bulk flow is typically defined as the average velocity within a volume $V$ of radius $R$ centred on the observer, assuming a top-hat filter~\citep{Watkins_2009,Feldman_2010,Nusser_2011,
Hoffman_2015,Watkins_2015,Nusser_2016,Scrimgeour_2016,Feix_2017,Hellwing_2018,Peery_2018,Whitford_2023,Watkins_2023}. In our case, three possible contributions enter:
\begin{equation}\label{eq:bulk_flow_components}
\begin{split}
    \bm{B}(R) = \frac{1}{V} \int_{V} \bigg[
        & \beta^\star \bm{v}(\bm{r}) + \Vext(\bm{r}) \\
        & - H_0 r \left(10^{\Delta_{\rm ZP}/5} - 1\right)
          \frac{\bm{\Delta}_{\rm ZP}}{\Delta_{\rm ZP}}
    \bigg] \dd V ,
\end{split}
\end{equation}
where the first term is the velocity field of~\citetalias{Carrick_2015} scaled by $\beta^\star$, the second is the external bulk flow $\Vext$ (optionally position dependent), and the third is the bulk-flow equivalent of the zero-point dipole (or equivalently $H_0$ dipole), here approximated with a linear Hubble law (the full expression yields negligible differences). We assume here $H_0 = 73.04~\kmsecMpc$~\citep{Riess_2022} for illustrative purposes: the exact results depend minimally on the assumed $H_0$. Sources along the zero-point dipole direction are intrinsically fainter (i.e. have a larger zero-point), inducing a dipole in their apparent magnitudes but not in their redshifts, since they are assumed to be uniformly distributed in volume. If the model accounts only for a velocity dipole and not the zero-point dipole, it assigns all sources the same zero-point and hence absolute magnitude on average. Consequently, sources along the dipole direction are inferred as more distant. To match the isotropy of redshifts, the model compensates by introducing a negative velocity dipole along the zero-point direction, hence the minus sign in~\cref{eq:bulk_flow_components}.

We compare the magnitude of this bulk flow to the \ac{LCDM} expectation, computed with a top-hat filter of radius $R$ (see e.g. Sec.~6.1.2 of~\citealt{Boruah_2019}), assuming \textit{Planck} 2018 cosmology~\citep{Planck_2020_cosmo}. This comparison is essential: while the data may show a strong preference for a dipole model (or more generally the existence of a bulk flow), it could still be entirely consistent with \ac{LCDM}.

\begin{figure}
    \centering
    \includegraphics[width=\columnwidth]{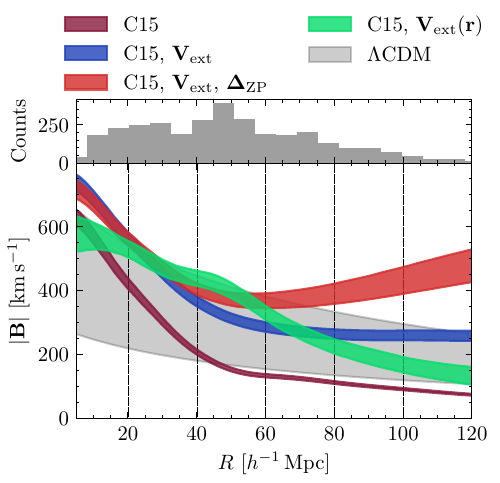}
    \caption{Bulk flow as a function of radius $R$ inferred from the \ac{CF4} W1 sample. We first show the bulk flow from~\protect\cite{Carrick_2015} only and then introduce three extensions of it: (i) a constant external flow $\Vext$, (ii) a constant $\Vext$ combined with the zero-point dipole $\bm{\Delta}_{\rm ZP}$, and (iii) a radially varying $\Vext(R)$. For reference, the shaded band shows the \ac{LCDM} expectation computed with a top-hat filter of radius $R$, and the top panel displays the distribution of redshift distances in the \ac{CF4} W1 sample. All bands show the range between the $16$\textsuperscript{th} and 84\textsuperscript{th} percentile.
    (The width of the~\protect\citetalias{Carrick_2015} is due to the $\beta^\star$ posterior.) All cases are consistent with \ac{LCDM} expectations except the zero-point dipole model at large radii, where the (extrapolated) linearly growing dipole dominates. This is however disfavoured relative to the more flexible radially varying $\Vext$ model. For $\Vext(\bm{r})$, we place spline knots at $0,\,20,\,40,\,60,\,80$ and $100~\Mpch$, indicated by vertical dashed lines.}
    \label{fig:CF4_bulk_flow}
\end{figure}

We show the resulting bulk flow in~\cref{fig:CF4_bulk_flow} for three extensions of~\citetalias{Carrick_2015}: (i) a constant $\Vext$ as in our main analysis, (ii) a constant $\Vext$ combined with $\bm{\Delta}_{\rm ZP}$, and (iii) a radially varying $\Vext(R)$. (In all cases the velocities contained within~\citetalias{Carrick_2015} are included.) As expected, the first case asymptotes to a constant above $R \approx 60~\Mpch$, where the constant $\Vext$ dominates over the decaying contribution from~\citetalias{Carrick_2015}. The second case behaves similarly at small radii, since the zero-point dipole contribution scales with distance, but yields a larger bulk flow above $R \approx 40~\Mpch$ and then grows linearly once the dipole contribution dominates. Finally, the radially varying $\Vext(R)$ produces a bulk flow slightly lower than the constant $\Vext$ case at small radii, higher at $R \approx 40$--$60~\Mpch$, and decaying again at larger distances.

\begin{figure}
    \centering
    \includegraphics[width=\columnwidth]{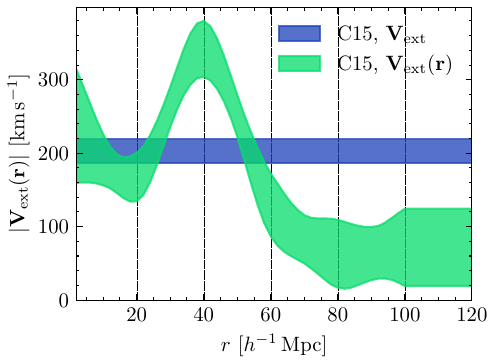}
    \caption{Magnitude of $\Vext(\bm{r})$ that is superimposed on the~\protect\citetalias{Carrick_2015} velocity field for a radially varying external flow, compared to the constant $\Vext$ case. The shaded band indicates the 16\textsuperscript{th} to 84\textsuperscript{th} percentile range. The magnitude rises and then decays towards zero, inconsistent with the linearly growing bulk flow expected from $H_0$ anisotropy. For $\Vext(\bm{r})$, we place spline knots at $0,\,20,\,40,\,60,\,80$ and $100~\Mpch$, indicated by vertical dashed lines. Above $100~\Mpch$ the model uses constant extrapolation.}
    \label{fig:CF4_Vext}
\end{figure}

In~\cref{fig:CF4_Vext} we show the magnitude of $\Vext(\bm{r})$ that is superimposed on the~\citetalias{Carrick_2015} velocity field. The magnitude rises from $200~\kmsec$ at small radii to about $300~\kmsec$ at $50~\Mpch$, after which it rapidly decays. $\Vext$ is interpolated between knots at $0,\,20,\,40,\,60,\,80$ and $100~\Mpch$, with the magnitude at each knot sampled uniformly between $0$ and $500~\kmsec$.~\cref{fig:CF4_Vext} demonstrates that this more flexible model is inconsistent with a linearly growing bulk flow, which would signal $H_0$ anisotropy. Furthermore, above $80~\Mpch$ inferred $\Vext$ is nearly consistent with zero, while the uncertainty band broadens largely due to the lack of data and the posterior transitioning to the prior. The posterior on the magnitude does not appreciably broaden beyond $100~\Mpch$, as sufficient data remain to constrain it at that distance; beyond this, we assume a constant extrapolation. The $\Vext(\bm{r})$ model has higher evidence than the $\Vext + \bm{\Delta}_{\rm ZP}$ model, and we verify that these conclusions are robust to the interpolation scheme and number and locations of the knots.

\section{Discussion and Conclusion}\label{sec:discussion_conclusion}

We revisited recent claims of Hubble constant anisotropy from direct distance tracers, made with the \ac{CF4} \ac{TFR} W1 subsample by~\citetalias{Boubel_2025}, and extended the analysis to the 2MTF and \SFI\ \ac{TFR} samples, as well as to the \PP\ and \PPL\ \ac{SN} samples (the latter being a reanalysis of \PP\ by~\citealt{Lane_2024}), all restricted to $z < 0.05$. Using a forward-modelling framework that simultaneously fits the distance relation, marginalises over distances, and models peculiar velocities, we tested for a dipole in the zero-point of the \ac{TFR} and in the standardised absolute magnitude of \acp{SN}. Since we do not apply an absolute distance calibration, this inferred zero-point represents a degenerate combination of the underlying distance relation zero-point and the Hubble constant.

\subsection{Summary of results}

In the \ac{CF4} W1 \ac{TFR} sample we infer a zero-point dipole amplitude of $0.087 \pm 0.019$ mag, with more than $2\sigma$ preference for a non-zero dipole magnitude in \SFI, \PP, and \PPL\ as well. Only in the 2MTF sample, which is also the least constraining, do we not infer a significant zero-point dipole. Under the assumption of the underlying zero-point being isotropic and the dipole arising from variation in $H_0$ alone (see Eq.~\ref{eq:H0_dipole}), the \ac{CF4} result corresponds to anisotropies in $H_0$ of approximately 4.1 per cent; see \cref{tab:dipole_summary} for a summary of the main results. We have confirmed that these results are robust to variations in the Galactic extinction correction. Specifically, we tested both the original $\ebv$ values used in \ac{CF4}~\citep{Schlegel_1998} and the maps of~\citet{Chiang_2023} or~\citet{Planck_2016}, jointly sampling the extinction coefficient $R_{\rm W1}$ in each case.

In the \ac{CF4} sample, Bayesian evidence ratios (with a uniform prior on dipole magnitude and isotropic orientation) strongly favour the anisotropic model, with odds of $877\!:\!1$. The \SFI\ sample provides only very weak support for anisotropy ($3\!:\!1$), while the 2MTF sample very weakly prefers isotropy ($2\!:\!1$). For the two \PP\ variants, the magnitude covariance matrix prevents a straightforward evidence computation, but the dipole magnitude is more than $3.5\sigma$ discrepant from zero.

Such a dipole in $H_0$ can equivalently be interpreted as a bulk flow that grows linearly with distance. To test if this is the preferred radial dependence we introduce a non-parametric model of radially varying $\Vext$ on top of the~\citetalias{Carrick_2015} field. We find that its magnitude increases roughly linearly up to $50~\Mpch$ but then rapidly decays towards zero (see~\cref{fig:CF4_Vext}), inconsistent with any $H_0$ anisotropy. The inferred bulk flow from this radially varying $\Vext$ remains consistent with \ac{LCDM} at all radii, decaying with $r$ as expected (see~\cref{fig:CF4_bulk_flow}). Thus, while the data favours an anisotropic model over one with a constant $\Vext$, a non-linearly-rising bulk flow is preferred which does not challenge \ac{LCDM}. The preference for a dipole in $H_0$ is caused by the relative inflexibility of that model and limited radial range of the data.

Moreover, we caution that beyond the Galactic extinction already considered, other angular selection effects could mimic a dipole-like signal. Since our analysis constrains only a degenerate combination of the dipole in the distance-tracer zero-point and the Hubble constant, the detected dipole could in principle reflect the former rather than a cosmological origin. To mitigate this, we tested several independent datasets with different calibrations and found consistent hints of a dipole in both the \ac{TFR} and \acp{SN}. For instance, in the case of the \ac{CF4} W1 sample,~\citetalias{Boubel_2025} found no evidence of anisotropy in the linewidths, while the use of consistent WISE photometry argues against spurious calibration dipoles. Nevertheless, because the \ac{CP} is a foundational principle of the standard cosmological model, extraordinary evidence is required for a robust detection, warranting further tests with future data.

Using \ac{CF4}-like mock catalogues, we showed that a dipole of amplitude $0.087$ (the posterior mean from our \ac{CF4} inference) is decisively favoured ($\log \mathcal{B} > 2$) once the sample size exceeds $\sim\num{2000}$. Smaller (mock) samples lack the statistical power to distinguish the models, consistent with the absence of decisive evidence in the 2MTF and \SFI\ samples.
By contrast, after quality cuts the \ac{CF4} W1 sample contains $\sim\num{3200}$ galaxies, for which the evidence strongly favours the anisotropic model, in line with the mock expectations. As shown in~\cref{fig:mock_evidences}, the evidence in favour of the anisotropic model increases steeply with sample size, highlighting the need for larger samples of well-calibrated distance tracers to achieve higher-significance detections. Extending to higher redshifts provides one route: peculiar velocities become subdominant at larger distances, reducing contamination from local flows. This gain, however, is offset by increased absolute distance uncertainties—typically $\sim20$ per cent for the \ac{TFR} and fundamental plane, and $\sim10$ per cent for Type Ia \acp{SN}. Our work highlights however that one must be extremely careful in claiming evidence for a fundamental anisotropy, because this may seem to be preferred even if the data really only contains a bulk flow consistent with \ac{LCDM}. Higher redshift data is crucial for breaking this degeneracy because the \ac{LCDM} bulk flow goes to zero at large $r$ while an anisotropic $H_0$ causes an effective bulk flow that continues to rise.

\subsection{Comparison to previous work}

We compare our results for the \ac{CF4} W1 \ac{TFR} sample with those of~\citetalias{Boubel_2025}, who also employ the~\citetalias{Carrick_2015} reconstruction to account for peculiar velocities and report a dipole of magnitude $0.063 \pm 0.016$ mag in the direction of $(\ell, b) = (142 \pm 30^\circ, 52 \pm 10^\circ)$, in good agreement with our measurement, though slightly lower and with smaller uncertainty. They find only weak support for the anisotropic model, reporting odds of $3\!:\!1$ in favour of the zero-point dipole. The discrepancy in the dipole magnitude likely stems from a combination of the following factors:
\begin{enumerate}
    \item \emph{Velocity assignment}. While our model queries the velocity field in real space, assigning velocities based on inferred real-space distances,~\citetalias{Boubel_2025} work in redshift space, first mapping the~\citetalias{Carrick_2015} field into redshift space using the prescription of~\citet{Carr_2022}, and then assigning peculiar velocities as a function of the observed redshift of each galaxy (while approximately undoing the fiducial $\bm{V}_{\rm ext}$ and $\beta^\star$ of~\citetalias{Carrick_2015} and applying new ones). The observed redshift together with the peculiar velocity then yields an estimate of the cosmological redshift with a scatter of $\sigma_v$. This approach introduces complications in the so-called triple-valued zones~\citep{StraussWillick_1995}, where a given observed redshift corresponds to multiple possible distances along the line of sight. In such regions, the mapping from redshift to peculiar velocity becomes non-unique, whereas the real-space assignment used in our framework retains a one-to-one correspondence between distance and peculiar velocity. \citetalias{Boubel_2025} use the observed data to infer latent model parameters (cosmological redshift), which are then used to predict the apparent magnitude. Our method more straightforwardly forward-models the observables and hence avoids these complications.
    \item \emph{Lower redshift cut applied to the W1 sample}. While~\citetalias{Boubel_2025} impose $c\zCMB > 3000~\kmsec$, we do not apply such a cut in our main analysis. We verify that applying the same cut yields a dipole magnitude of $0.062 \pm 0.022$ in the direction $(\ell, b) = (137 \pm 19^\circ, 24 \pm 12^\circ)$, both of which are in a better agreement with~\citetalias{Boubel_2025} than our fiducial result.
    \item \emph{Photometry-quality cut}. \citetalias{Boubel_2025} do not apply a photometry-quality cut, thus retaining approximately $\num{1300}$ more galaxies in the \ac{CF4} \ac{TFR} W1 sample. We test repeating our fiducial analysis without applying the quality cut and recover a dipole of magnitude of $0.066 \pm 0.018$ mag in the direction $(\ell, b) = (128 \pm 14^\circ, 18 \pm 10^\circ)$. This magnitude is in good agreement with~\citetalias{Boubel_2025}, though the direction remains slightly off.
\end{enumerate}
Thus, after matching the lower redshift or photometry-quality cut, we find a dipole in reasonable agreement with~\citetalias{Boubel_2025}, though the direction remains mildly inconsistent. The residual differences with~\citetalias{Boubel_2025} are most likely due to the velocity-assignment scheme.

Our work is part of the larger programme of searching for deviations from the \ac{CP} through anisotropy, which is receiving increasing attention. On the \ac{TFR} side, besides~\citetalias{Boubel_2025} the studies most directly comparable to ours examine bulk flows in the local Universe. These are strongly correlated with anisotropic Hubble expansion, and indeed if the bulk flow is linearly rising, then locally they are indistinguishable. Claims for such behaviour include~\citet{Watkins_2009,Watkins_2015,Watkins_2023}, the latter in particular using the same \ac{CF4} data as us and claiming a bulk flow of $\sim$$400~\kmsec$ at $250~\Mpch$, in $\sim$5$\sigma$ tension with cosmic variance in \ac{LCDM}. However, in contrast to~\cite{Watkins_2023}, we find a decaying bulk flow, consistent with \ac{LCDM} expectations.

In our analysis we infer a dipole in the degenerate quantity $\aTFR + 5 \log h$ (or $\MSN + 5 \log h$). Interpreted as a dipole in $H_0$, this implies that along the dipole direction $H_0$ is enhanced, requiring a negative peculiar velocity to reproduce the same observed redshift. Since the aforementioned studies report a dipole in peculiar velocity, we compare the opposite of their inferred directions with our result. Working in the \ac{CMB} frame, we compare our inferred dipole with literature estimates in~\cref{fig:dipole_comparison}. The dipole direction we find in \ac{CF4} agrees with both~\cite{Migkas_2021} and~\cite{Watkins_2023}, the latter using a larger set of \ac{CF4} data. By contrast, the \PP\ dipole direction disagrees with our \ac{TFR} sample but matches well with the \PP\ analysis of~\cite{Sorrenti_2023}, which may reflect the different redshift ranges probed.

On the \ac{SN} side, several studies find significant or semi-significant anisotropies, (e.g.~\citealt{Hu,SN_1,SN_2,Sah,Conville,Krishnan,Zhai,Rahman_2022,Cowell_2023,Sorrenti_2025}). \cite{Rahman_2022} investigated a possible dipole in the distance moduli of Type Ia \acp{SN} using the JLA sample~\citep{Betoule_2014}. They applied the reconstruction of~\citetalias{Carrick_2015} to correct for peculiar velocities and found no evidence for anisotropic expansion. However, other than this study, typically these \acp{SN} studies are at higher redshift than we consider; we have cut our sample to be consistent with the \ac{TFR} data and to remain within the volume covered by the~\citetalias{Carrick_2015} reconstruction ($z < 0.05$). In addition the methodologies differ: instead of fitting dipoles (or multipoles) some studies perform separate analyses in different parts of the sky, which reduces statistical constraining power but enables investigation of more general types of anisotropy. A limiting case of this is the ``hemispherical fitting'' method which compares results in one half of the sky to the other. Sometimes the fitting is done in frames other than the \ac{CMB} frame (e.g.~\citealt{Sah}), sometimes anisotropy specifically in the direction of the \ac{CMB} dipole is searched for (e.g.~\citealt{Zhai, Krishnan}), and sometimes more sophisticated frameworks are employed (e.g.~\citealt{Basheer,Heinesen}). The situation is therefore still unclear as to whether \acp{SN} evince a bona fide anisotropy (though we find evidence for it with the \PP\ samples). We caution however that such results are sensitive to the peculiar velocity field, which is not well known beyond $z\approx0.05$.

\subsection{Disentangling the cosmological dipole from coherent flows}

Accounting for peculiar velocities assuming the~\citetalias{Carrick_2015} reconstruction may (partially) absorb a potential cosmological $H_0$ dipole, making the disentanglement of a cosmological dipole in $H_0$ from coherent flows challenging. In principle, anisotropies in the expansion rate would imprint on the density field inferred from redshift-space positions, thereby biasing the reconstructed peculiar velocities. While the~\citetalias{Carrick_2015} reconstruction assumes linear theory and $\Lambda{\rm CDM}$ to relate the density and velocity fields via the continuity equation and growth rate, it does not constrain the amplitude or anisotropy of the density fluctuations beyond the data itself. Consequently, any non-standard large-scale anisotropy in expansion could propagate into the reconstructed velocity field and be inadvertently subtracted. Nevertheless, the measured clustering amplitude of the~\TWOMPP\ galaxies remains consistent with $\Lambda{\rm CDM}$ expectations. The present analysis therefore searches for any dipolar excess beyond that predicted by~\citetalias{Carrick_2015} based on the~\TWOMPP\ galaxy positions.

\vspace{2em}
\noindent To conclude, using both \ac{TFR} galaxy samples and Type Ia \ac{SN} samples, we find (apparent) evidence for anisotropy in the zero-point calibration of the distance-tracer relation. In the \ac{CF4} and \PP\ sample, the implied variation in the Hubble constant is at the level of $4.1 \pm 0.9$ and $2.3\pm0.06$ per cent, respectively. Only in the less constraining \ac{TFR} samples (2MTF and \SFI) do we find no preference between the isotropic and anisotropic model. These results are robust to variations in Galactic extinction but are sensitive to the treatment of peculiar velocities, which are non-negligible at $z < 0.05$. Without accounting for peculiar velocities, the \ac{CF4} dipole would be even larger, while the \PP\ remains largely unaffected. We have also tested the reanalysis of~\PP\ by~\cite{Lane_2024}, which yields a dipole larger by $1\sigma$ compared to the fiducial~\PP\ data, corresponding to a variation in the Hubble constant of $2.8 \pm 0.6$ per cent.

While the preference for the zero-point dipole could in principle indicate a genuine anisotropy in the Hubble constant, a more plausible explanation is that it reflects local peculiar velocities not captured by~\citetalias{Carrick_2015} (or systematics) rather than a true cosmological anisotropy. Allowing $\Vext$ to vary radially, we find its magnitude rises initially and then rapidly decays to zero, yielding a bulk flow fully consistent with \ac{LCDM}.

These results highlight the importance of continued tests of the local Hubble flow, while also underscoring the challenge of disentangling genuine cosmological anisotropies from local peculiar velocities. Properly accounting for peculiar velocities and systematic effects is crucial when using direct distance tracers in cosmological analyses. It is only through such close scrutiny with robust statistical methodology that deviations from the \ac{CP} may be reliably detected.

\section*{Data availability}

The~\cite{Carrick_2015} reconstruction is available at \url{https://cosmicflows.iap.fr}. The public \ac{CF4} data is available at \url{https://edd.ifa.hawaii.edu/dfirst.php}. The code underlying the article is available at \url{https://github.com/Richard-Sti/CANDEL} and all other data will be made available on reasonable request to the authors.

\section*{Acknowledgements}

We thank Indranil Banik, Deaglan J. Bartlett, Nicholas Choustikov, Julien Devriendt, Pedro Ferreira, Sebastian von Hausegger, Mike Hudson, Mohamed Rameez, Animesh Sah, Subir Sarkar and Adrianne Slyz for useful inputs and discussions. We thank Anthony Carr for providing a version of the \PP\ covariance matrix with the peculiar velocity contributions removed. We thank Jonathan Patterson for smoothly running the Glamdring Cluster hosted by the University of Oxford, where a part of the data processing was performed. This project has received funding from the European Research Council (ERC) under the European Union's Horizon 2020 research and innovation programme (grant agreement No 693024). The authors would like to acknowledge the use of the University of Oxford Advanced Research Computing (ARC) facility in carrying out this work.\footnote{\url{https://doi.org/10.5281/zenodo.22558}}

RS acknowledges financial support from STFC Grant No. ST/X508664/1, the Snell Exhibition of Balliol College, Oxford, and the CCA Pre-doctoral Program. HD is supported by a Royal Society University Research Fellowship (grant no. 211046).
GL acknowledges support from the Simons Foundation through the Simons Collaboration on ``Learning the Universe''.

\bibliographystyle{mnras}
\bibliography{ref}

\appendix

\section{Flow model of S25}\label{sec:flow_model_full}

\subsection{Tully--Fisher flow model}

In this section, we summarise the flow model used in this analysis, which is based on our previous work in~\citetalias{VF_olympics}. For a more detailed description, we refer the reader to that study. We employ a forward model that jointly calibrates the \ac{TFR} (and its dipole) while also calibrating the local Universe reconstruction (which accounts for smaller-scale peculiar velocity modes) and other model parameters. For each galaxy, we have observations of its redshift, $\zCMB$ (converted to the \ac{CMB} frame) along with its uncertainty, $\sigma_{z}$. We define its value multiplied by the speed of light as $\sigma_{\rm cz} \equiv c \sigma_{\rm z}$. The total redshift of a galaxy in the \ac{CMB} frame is given by
\begin{equation}\label{eq:redshift_composition}
    1 + \zCMB = \left(1 + \zcosmo\right)\left(1 + \zpec\right),
\end{equation}
where $\zcosmo$ is the redshift due to cosmic expansion, and $\zpec = \Vpec /c$ represents the redshift contribution from the galaxy's line-of-sight peculiar velocity, $\Vpec$. In a flat \ac{LCDM} universe dominated by non-relativistic matter and dark energy, the cosmological redshift, $\zcosmo$, is related to the comoving distance, $r$, by (e.g.,~\citealt{Hogg1999})
\begin{equation}\label{eq:redshift_to_distance}
    r(\zcosmo) = \frac{c}{H_{\rm 0}} \int_{0}^{\zcosmo} \frac{\dd z^\prime}{\sqrt{\Om (1 + z^\prime)^3 + 1 - \Om}},
\end{equation}
where $\Om$ is the matter density parameter. The velocity field, $\bm{v}(\bm{r})$, is modelled under the single-flow approximation, so that the line-of-sight peculiar velocity is
\begin{equation}
    \Vpec = \left(\beta \bm{v}(\bm{r}) + \bm{V}_{\rm ext}\right) \cdot \hat{\bm{r}},
\end{equation}
where $\hat{\bm{r}}$ is the galaxy line-of-sight unit vector, and $\bm{V}_{\rm ext}$ accounts for external flows originating beyond the reconstruction volume. The parameter $\beta^\star$ is a calibration factor to scale the velocities predicted by~\citetalias{Carrick_2015} and is a function of both cosmology and the galaxy bias. When assuming no underlying velocity field, we effectively set $\bm{v} = 0$, modelling the flow solely as $\bm{V}_{\rm ext}$. A third parameter, $\sigma_v$, which we introduce later, captures small-scale velocity dispersion not accounted for by the reconstruction.

For each galaxy we observe the apparent magnitude $\mobs$ with uncertainty $\sigma_{\rm m}$ and the linewidth $\etaobs$ with uncertainty $\sigma_{\eta}$ (which is a distance-independent observable). These quantities constrain the galaxy distance. In \citetalias{VF_olympics} we introduce two latent parameters per galaxy to be inferred: the distance $r$ and the true linewidth $\etatrue$. The \ac{TFR} relates $\etatrue$ to the absolute magnitude $M$ via~\cref{eq:TFR_absmag}. Converting $r$ to a distance modulus $\mu$ yields a predicted apparent magnitude, $\mpred = \mu(r) + M(\etatrue)$, with the distance modulus being
\begin{equation}
    \mu = 5 \log \frac{d_{\rm L}}{\mathrm{Mpc}} + 25,
\end{equation}
where the luminosity distance $d_{\rm L}$ is related to the comoving distance $r$ by
\begin{equation}
    d_{\rm L} = (1 + \zcosmo) r,
\end{equation}
assuming a flat \ac{LCDM} universe.

We jointly infer the velocity field calibration parameters, ($\bm{V}_{\rm ext}, \beta, \sigma_v$), the distance indicator parameters, ($\aTFR, \bTFR, \cTFR, \sint$), the mean $\hat{\eta}$ and standard deviation $w_\eta$ of the $\etatrue$ prior, and potentially the zero-point dipole $\bm{\Delta}_{\rm ZP}$. Assuming independent sources,~\citetalias{VF_olympics} formulate the likelihood of the observed redshift, magnitude, and linewidth given the distance, true linewidth, and the set of model parameters $\bm{\theta}$ as
\begin{equation}\label{eq:likelihood}
    \begin{split}
        \mathcal{L}(\zobs,\,\mobs,\,&\etaobs \mid \bm{\theta},\,r,\,\etatrue)
        =\\
        &=\mathcal{N}\left(c\zobs; c\zpred,\,\sqrt{\sigma_v^2 + \sigma_{c\zobs}^2}\right)\\
        &\times\mathcal{N}\left(\mobs; \mpred,\,\sqrt{\sint^2 + \sigma_{m}^2}\right)\\
        &\times\frac{\mathcal{N}(\etaobs; \etatrue,\,\sigma_\eta)}{p(S = 1 \mid \hat{\eta},\,w_\eta)},
    \end{split}
\end{equation}
where $p(S = 1 \mid \hat{\eta},\,w_\eta)$ is the expected fraction of retained sources given a truncation in $\etaobs$. Strictly speaking, $p(S = 1 \mid \hat{\eta},\,w_\eta)$ is not part of the data likelihood. It should multiply the product of the per-sample likelihoods and the prior as $\left[p(S = 1 \mid \hat{\eta},\,w_\eta)\right]^{-n}$, where $n$ is the number of galaxies in the sample. For notational convenience, however, we absorb it into the likelihood (see~\citealt{Kelly_2008} for more details). The first term in~\cref{eq:likelihood} is the likelihood of the observed redshift given the predicted value, which depends on the inferred distance and peculiar velocity, while the second term is the likelihood of the observed magnitude given the predicted apparent magnitude. Furthermore, we have that
\begin{equation}
\begin{split}
    p(S = 1 \mid \hat{\eta},\,w_\eta)
    &=
    \iint \mathrm{d}\eta_{\rm obs}\,\mathrm{d}\eta_{\rm true}\;
    p(S = 1 \mid \eta_{\rm obs}) \\
    &\quad\times \mathcal{L}(\eta_{\rm obs} \mid \eta_{\rm true})\,
    \pi(\eta_{\rm true} \mid \hat{\eta},\,w_\eta),
\end{split}
\end{equation}
where $p(S = 1 \mid \eta_{\rm obs})$ is a binary detection indicator between $\eta_{\min}$ and $\eta_{\max}$,
\begin{equation}
    p(S = 1 \mid \eta_{\rm obs})
    =
    \begin{cases}
        1 & \text{if}\quad \eta_{\min} < \eta_{\rm obs} < \eta_{\max}, \\[3pt]
        0 & \text{otherwise}.
    \end{cases}
\end{equation}
$\mathcal{L}(\eta_{\rm obs} \mid \eta_{\rm true})$ denotes the Gaussian likelihood of the observed given the true linewidth, and $\pi(\eta_{\rm true} \mid \hat{\eta},\,w_\eta)$ is the Gaussian prior on the true linewidth. Given these assumptions, it can be shown that
\begin{equation}
    p(S = 1 \mid \eta_{\rm obs}) = \Phi\!\left(\frac{\eta_{\max} - \hat{\eta}}
                      {\sqrt{\sigma_\eta^{2} + w_\eta^{2}}}\right)
    -
    \Phi\!\left(\frac{\eta_{\min} - \hat{\eta}}
                      {\sqrt{\sigma_\eta^{2} + w_\eta^{2}}}\right),
\end{equation}
where $\Phi(x)$ is the cumulative density function of the standard normal distribution, defined as
\begin{equation}\label{eq:CDF_standard_normal}
    \Phi(x) = \frac{1}{\sqrt{2\pi}} \int_{-\infty}^x e^{-t^2/2} \dd t.
\end{equation}
We assume only a lower threshold in $\etaobs$ in \ac{CF4} and \SFI, while in 2MTF we impose both a lower and upper threshold. We marginalise over $r$ and $\eta_{\rm true}$ as
\begin{equation}
    \begin{split}
        \mathcal{L}(\zobs,\,&\mobs,\,\eta_{\rm obs} \mid \bm{\theta})
        =\\
        &
        =\iint \mathcal{L}(\zobs,\,\mobs,\,\eta_{\rm obs} \mid \bm{\theta},\,r,\,\eta_{\rm true}) \\
        &\quad\times \pi(r \mid \bm{\theta})\,\pi(\eta_{\rm true} \mid \bm{\theta})\,\mathrm{d}r\,\mathrm{d}\eta_{\rm true},
    \end{split}
\end{equation}
where $\pi(r \mid \bm{\theta})$ is defined in~\cref{eq:empirical_prior_distance}.
Rather than modelling the full H\textsc{I} selection of the \ac{TFR} sample, which is the primary reason why the \ac{TFR} samples do not extend to higher redshifts, we follow~\cite{Lavaux_Virbius} and adopt an effective treatment by setting the distance prior to~\cref{eq:empirical_prior_distance}, which incorporates both the homogeneous and inhomogeneous Malmquist bias. We may either sample $r$ and $\eta_{\rm true}$ for each galaxy directly with an \ac{HMC} sampler, or marginalise over them numerically at each \ac{MCMC} step. The former is computationally faster, while the latter yields a lower-dimensional parameter space suitable for evidence computation. Because evidence values are central to this work, we adopt the latter approach and evaluate the two-dimensional integral numerically on a grid in $r$ and $\eta_{\rm true}$ at each \ac{MCMC} step. We define a fixed radial distance grid ranging from $0.001$ to $201~\Mpch$ with a step size of $0.5~\Mpch$, which is sufficient given that the~\citetalias{Carrick_2015} field is smoothed on scales of $4~\Mpch$. For $\etatrue$, we adopt an adaptive binning scheme. The Gaussian likelihood term $\mathcal{L}(\etaobs \mid \etatrue)$ determines the region of $\etatrue$ that carries non-negligible probability mass. Accordingly, for each source we define a grid spanning $\etaobs \pm 5\sigma_\eta$, discretised into $31$ equally spaced steps, and evaluate the likelihood over a $402\times31$ grid at each \ac{MCMC} iteration. We then marginalise over this two-dimensional grid using Simpson's rule. This computation is performed on GPUs using our \texttt{JAX}-based implementation\footnote{\url{https://github.com/jax-ml/jax}}. To verify that the numerical integration does not introduce bias, we compared the results to a model in which the latent parameters $\etatrue$ were explicitly sampled and found the posteriors to be consistent.

We compute the model evidence, $\mathcal{Z}$, defined as the integral of the likelihood weighted by the prior over the parameter space,
\begin{equation}\label{eq:evidence_definition}
    \mathcal{Z} \equiv \int \dd \bm{\theta} \mathcal{L}(D \mid \bm{\theta}) \pi(\bm{\theta}),
\end{equation}
where $D$ represents the data and $\bm{\theta}$ the model parameters. The ratio of evidences between two models, known as the Bayes factor, quantifies the relative statistical support for one model over another, assuming equal prior model probabilities.

\section{CF4 TFR W1 full posterior}\label{sec:full_posterior}

\begin{figure*}
    \centering
    \includegraphics[width=\textwidth]{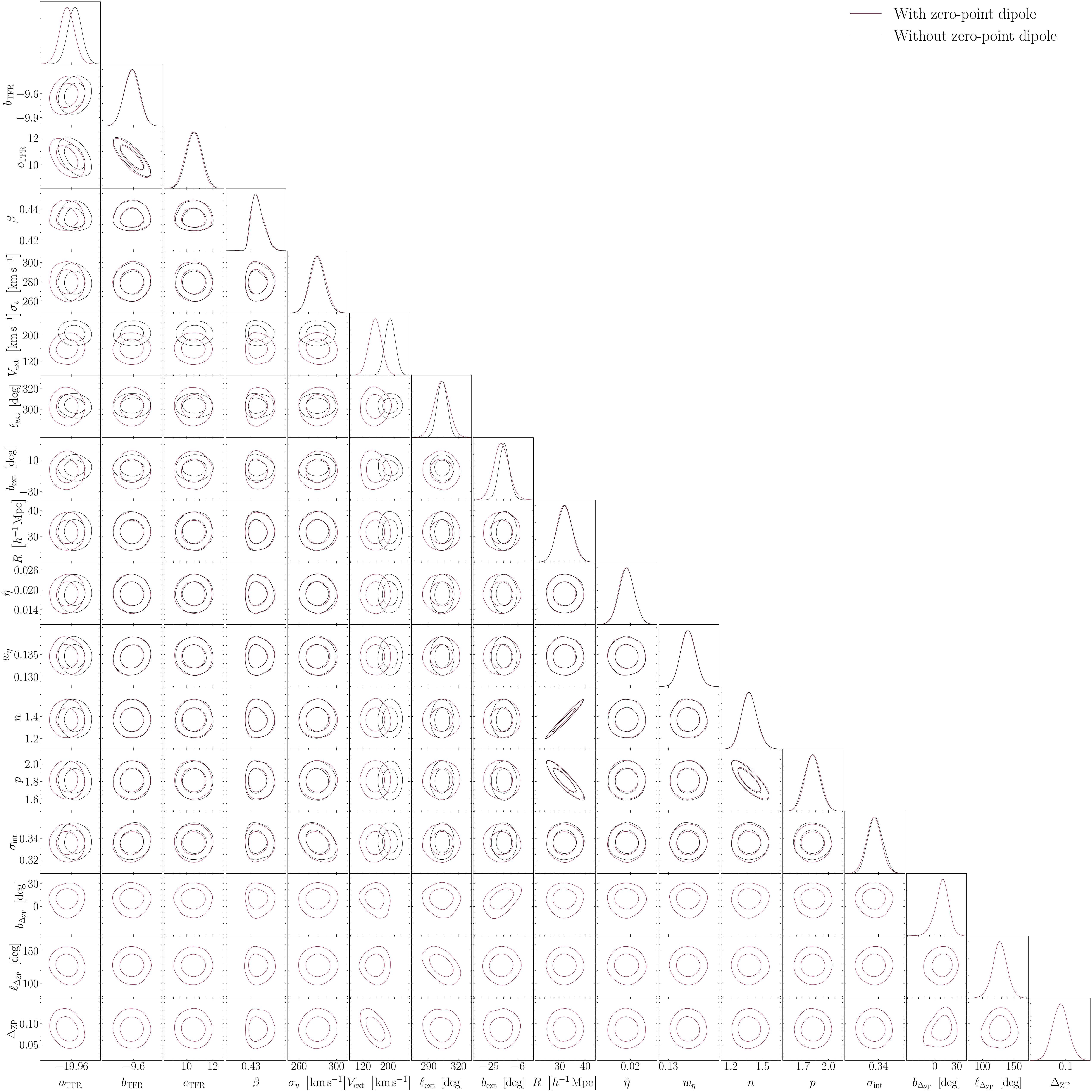}
    \caption{Posterior distributions for the model parameters inferred from the \ac{CF4} \ac{TFR} W1 data,  comparing the zero-point dipole model (red) to the isotropic model (black). Parameters shown include the \ac{TFR} calibration ($\aTFR$, $\bTFR$, $\cTFR$), velocity field calibration factor ($\beta^\star$), scatter terms ($\sint$, $\sigma_v$), external flow parameters ($V_{\rm ext}$, $\ell_{\rm ext}$, $b_{\rm ext}$), linewidth hyperparameters ($\hat{\eta}$, $w_\eta$), distance prior hyperparameters ($R$, $n$, $p$), and zero-point dipole parameters ($\Delta_{\rm ZP}$, $\ell_{\Delta_{\rm ZP}}$, $b_{\Delta_{\rm ZP}}$). We observe a mild degeneracy between the magnitude of the zero-point dipole and the external flow, and find that introducing a dipole shifts the inferred $\Vext$ while leaving other parameters largely unchanged. Contours denote $1\sigma$ and $2\sigma$ credible regions.}
    \label{fig:C15_posterior_CF4_W1}
\end{figure*}

In~\cref{fig:C15_posterior_CF4_W1}, we present the posterior distribution over model parameters inferred from the \ac{CF4} \ac{TFR} W1 data, with and without a dipole in the zero-point. The parameter set includes the isotropic \ac{TFR} zero-point, slope, and curvature ($\aTFR$, $\bTFR$, $\cTFR$); the velocity field calibration factor $\beta^\star$; the intrinsic \ac{TFR} scatter ($\sint$); the redshift scatter ($\sigma_v$); the external flow parameters ($V_{\rm ext}$, $\ell_{\rm ext}$, $b_{\rm ext}$); the zero-point dipole parameters ($\Delta_{\rm ZP}$, $\ell_{\Delta_{\rm ZP}}$, $b_{\Delta_{\rm ZP}}$); the linewidth hyperprior mean and standard deviation ($\hat{\eta}$, $w_\eta$); and the distance prior parameters ($R$, $n$, $p$).

We find a mild degeneracy between the zero-point dipole amplitude and the magnitude of the external flow: a larger external flow can compensate for a smaller zero-point dipole, and vice versa. Comparing the model with a zero-point dipole to the isotropic case, we find a slight shift in the zero-point monopole, while the posteriors for $\bTFR$, $\cTFR$, $\alpha$, $\beta^\star$, $\sint$, $\hat{\eta}$, and $w_\eta$ remain unchanged. As expected, introducing the dipole shifts the posterior on $\Vext$, affecting both its magnitude and Galactic latitude.

\bsp
\label{lastpage}
\end{document}